\documentclass{jaa}
\usepackage{natbib}
\usepackage{url}

\usepackage{graphicx, subfigure }

\begin{document}\sloppy

%%paper title
%%For line breaks \\ can be used within title
\title{Orbits and vertical height distribution of 4006 open clusters in the Galactic disk using Gaia DR3\\}

%%author names are separated by comma (,)
%%use \and before the last author name
%%use a * along with the number separated by comma
%% for the  author for correspondence
%%\textsuperscript{number} is used for affiliation
%%\affilOne, \affilTwo etc., upto \affilTwentyfive is possible
%%Please note the first letter after \affil is capitalised in the command
%%

\author{Geeta Rangwal\textsuperscript{1,2,*}, Aman Arya\textsuperscript{3,4,*}, Annapurni Subramaniam\textsuperscript{2}, Kulinder Pal Singh\textsuperscript{4} and Xiaowei Liu\textsuperscript{1}}
\affilOne{\textsuperscript{1}South-Western Institute for Astronomy Research,
 Yunnan University, Kunming 650500, People's Republic of China\\}
\affilTwo{\textsuperscript{2}Indian Institute of Astrophysics, Bangalore, India\\}
\affilThree{\textsuperscript{3}Stony Brook University, Stony Brook, New York 11790, USA\\}
\affilFour{\textsuperscript{4}Department of Physical Sciences, Indian Institute of Science Education
Research (IISER)- Mohali, SAS Nagar, Punjab 140306, India\\}
%%escape two column mode for title, affiliation and abstract
%%by giving \twocolumn command as shown

\twocolumn[{

\maketitle

%%include \corres to print the corresponding author Email id
\corres{geetarangwal91@gmail.com, amanarya1910@gmail.com}

%%include \msinfo for
%%manuscript information such as
%%received, revised and accepted dates
%%
% \msinfo{1 January 2015}{1 January 2015}

%%abstract
\begin{abstract}

Open clusters (OCs) in the Galaxy are excellent probes for tracing the structure and evolution of the Galactic disk. 
We present an updated catalog of the fundamental and kinematic parameters for 1145 OCs, estimated using the data from Gaia DR3 earlier listed in \citet{2020A&A...640A...1C}.
This sample is complemented by 3677 OCs with astrometric solution from the catalogue by \citet{2023A&A...673A.114H}.
Using the Galaxy potential and the space velocities, orbits of 4006 OCs were computed, and we provide a catalogue with orbital parameters such as eccentricity, perigalactic and apogalactic distance, and the maximum vertical height traced by OCs from the Galactic disk. The OCs in the sample are found to be distributed between 5-16 kpc from the Galactic center, with older OCs showing a radially extended distribution.
The low number of old OCs in the inner region of the Solar circle is likely to suggest their destruction in this area.
We explored the maximum vertical height ($Z_{max}$) OCs can reach using the orbital estimations. We derive a quantitative expression for the dependency of $Z_{max}$ with the cluster's age and Galactocentric radius for the first time. The young (age $<$ 50 Myr) and the intermediate age ( 50 Myr $<$age $<$ 1 Gyr) OCs show similar values of $Z_{max}$ till 9 kpc, with the latter group higher values beyond.
OCs older than 1 Gyr show larger values of $Z_{max}$ at all Galactocentric radii and significantly larger values beyond 9 kpc.
Higher values of $Z_{max}$ are found in the third Galactic quadrant, suggesting the link between the higher values and the Galactic warp.
This large sample shows that young OCs are also involved in the diagonal ridge formation in the solar neighborhood.

\end{abstract}

%%insert keywords separated by 3 hyphens using \keywords{words}
}]

\doinum{12.3456/s78910-011-012-3}
\artcitid{\#\#\#\#}
\volnum{000}
\year{0000}
\pgrange{1--}
\setcounter{page}{1}
\lp{1}

%%%%%%%%%%%%%%%%% BODY OF PAPER %%%%%%%%%%%%%%%%%%

\section{Introduction}\label{sec:intro}

Our Galaxy, the Milky Way, has three easily distinguishable major components: bulge, halo, 
and disk. Again, the Galaxy disk is made up of thin and thick disks.
Due to our position in the disk of the Galaxy, it isn't easy to study the
a global picture of the Galaxy, but the precise observations of position and 
the motion of the stars in the sky by the Gaia mission is
transforming our understanding of the Galaxy \citep{2023A&A...674A...1G}.
For example, the studies by \citet{2018Natur.563...85H, 2018MNRAS.478..611B, 2018MNRAS.481.3442M, 2019MNRAS.488.1235M, 2020ARA&A..58..205H, 2020MNRAS.497L...7S, 2021ApJ...907L..16R}  showed that the different components of the Galaxy, such as thin disk, thick
disk, bulge, and halo are interlinked formation phases of a system that were previously assumed to be distinct.

The well-established phenomenon of star formation states that stars form in
groups due to the gravitational collapse of progenitor molecular cloud \citep{lada2003embedded}, resulting
in star clusters. Open star clusters (OCs) in the Galaxy are found in a range of age
and mass and distributed at various $R$ in the Galactic disk, including thin and thick disks.
Determination of the fundamental properties like age, distance, and proper motion
is more precise for star clusters than for individual stars in the disk.
Which made them wonderful laboratories for studying stellar dynamics
\citep{1995ARA&A..33..381F} and
the structure of the Galactic disk.

Multiple studies have compiled catalogues of OC parameters (age, distance, and extinction) by combining parallaxes with photometric data from Gaia data releases. \cite{2018A&A...615A..49C} obtained mean astrometric parameters (proper motions and parallaxes) for 128 OCs closer than about 2 kpc from the Sun by applying an unsupervised membership assignment procedure to select high probability cluster members from the full astrometric dataset from the first release of Gaia data \citep{2016A&A...595A...1G} and Tycho-2 \citep{2016A&A...595A...4L}. \cite{2018A&A...618A..93C}, in their study, obtained a list of 1229 OCs from Gaia data alone and derived their mean parameters (particularly distances) by applying an unsupervised membership assignment code (\textit{UPMASK}) trained on data from thousands of clusters recorded previously in the literature. \cite{2019A&A...623A.108B} applied an automated Bayesian tool, BASE-9, to fit stellar isochrones on the observed $G$, $G_{BP}$, $G_{RP}$ magnitudes of the high probability member stars (solely from Gaia). Selecting only the low-reddening objects and discarding the very young OCs (for which isochrone fits do not give the best age estimates), they primarily determined the age of 269 OCs. 

\cite{2021MNRAS.504..356D} used published membership probability of stars derived from Gaia DR2 data, applied isochrone fitting to Gaia data, and estimated astrophysical parameters for 1743 OCs. Furthermore, they published mean radial velocity values for 831 OCs, of which 198 were new. 
\cite{2021A&A...652A.102H} considered the list of OC members published from the pre-Gaia era (hence not as precise measurements as from Gaia) and compiled a heterogenous catalogue of 3794 OCs.
The largest, most homogenous, and most accurate catalogue of OCs, compiled solely from Gaia data DR2, was published by \cite{2020A&A...640A...1C}. 
They used an artificial neural network trained over objects in the literature from the pre-Gaia era and estimated the distance and age of 2017 OCs. Of these 2017 clusters, only 1867 clusters with reliable parameters were used. The remaining 150 OCs were found to either have blurred CMDs or 
have very few member stars, which were insufficient to estimate their age and distance. Furthermore, these 1867 OCs only contain stars with a membership 
probability greater than 70 \%. With the availability of Gaia DR3, it is crucial to update the parameters of this sample of 1867 OCs. 

Recently, \citet{2023A&A...673A.114H} (hereafter, HR23) presented an updated catalogue of OCs in the Milky Way using the data from Gaia DR3.
They used the Hierarchical Density-Based Spatial Clustering of Applications with Noise (HDBSCAN) algorithm and recovered 7167 star clusters in total. 
These clusters are significant probes to study various aspects, including star formation history, stellar evolution, properties of the Galaxy, and its kinematics, 

In this study, we perform the parameter update of OCs in \cite{2020A&A...640A...1C}, then combine this with OCs of HR23 to derive the orbits of the OCs in the Galaxy.  Based on the orbital estimations, some structural properties of the Galactic disk, as delineated by the OCs, are presented as a function of age and Galactic radius. The data used for this study are given in section \ref{sec:data and catalogue}. The method adopted to derive the orbits is discussed in
section \ref{sec:orbits} and its results are illustrated in 
section \ref{sec:results}. We discussed our findings in the section \ref{sec:discussion} and summarised the results in \ref{sec:summary}.

\section{Data and Sample clusters}\label{sec:data and catalogue}
\subsection{Data}

The present study uses the kinematical data from the third installment of the dataset published by the European Space Agency's Gaia mission. 
Gaia is an astronomical survey designed to create a precise 3D Milky Way galaxy map, launched in December 2013 \citep{2016A&A...595A...1G}.
Gaia's third data release (DR3) \citep{2023A&A...674A...1G} has provided an unprecedented five-parameter astrometric 
solution - position on the sky ($\alpha, \delta$), parallax, and proper motion, and broadband photometry in $G$, $G_{BP}$, \& $G_{RP}$ passbands, 
for around 1.3 billion sources, with a limiting magnitude of G $\approx$ 21 and a bright limit of G $\approx$ 3. 
Parallax uncertainties range up to 0.7 mas towards the fainter end. Gaia DR3 has made significant leaps by providing the mean line of sight velocities for more than 33 million objects.
%%%%%%%%%%

\subsection{catalogues of open clusters in the Galaxy}

We aim to use data from Gaia DR3 to update the OC parameters in the catalogue by \citet{2020A&A...640A...1C} based on Gaia DR2. Since the same physical objects have different source-id in different data releases from Gaia, direct one-to-one correspondence between numerical identifiers from Gaia DR2 to Gaia DR3 is not guaranteed. However, Gaia provides a pre-computed cross-match table containing sources between Gaia DR2 and Gaia DR3. First, wherever possible, the sources included in \citet{2020A&A...640A...1C} are corrected for their linear proper motion from Gaia DR2 to Gaia DR3. Then, a cone search is performed for all objects, and the objects within the neighborhood of a 2" radius are selected as the best suitable cross-match pairs. The angular distance and magnitude difference are simultaneously calculated for every pair. 

 \begin{figure}
        \centering
        \includegraphics[width=0.8\columnwidth]{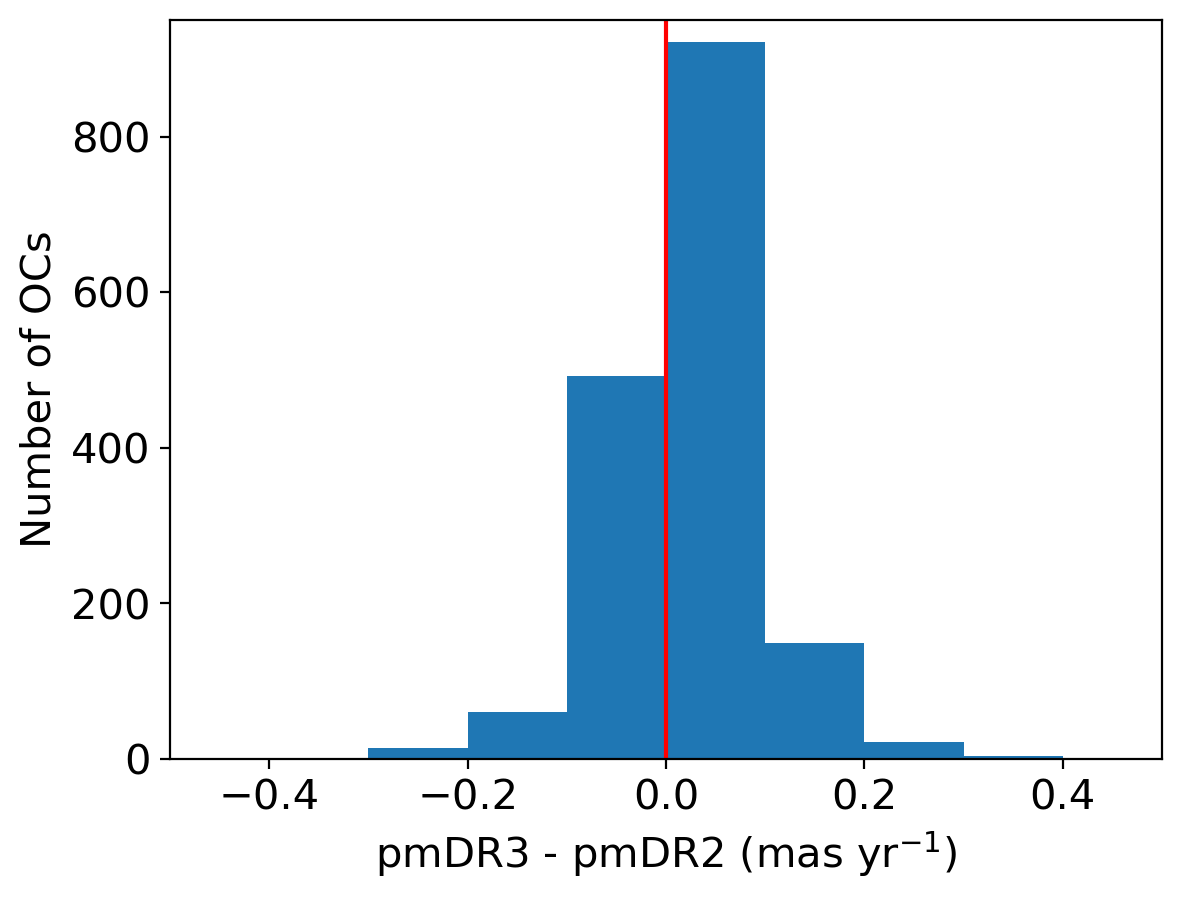}
        \qquad
        \caption{A histogram representing the shift in proper motion from Gaia DR2 to Gaia DR3 within the epoch of half a year.}
        \label{schematic}
\end{figure}

A more accurate proper motion propagation was required from Gaia DR2 to Gaia DR3. Still, the cone search with the same radius for all the stars gave comprehensive results in Gaia DR3 against Gaia DR2, even though there is a slight average epoch difference of half a year between the two catalogues. We note that only a small sample (less than 0.2\%) did not find their counterparts in Gaia DR3. Fig. \ref{schematic} shows the difference of mean proper motions of OCs calculated in the present analysis using Gaia DR3 and from \citet{2020A&A...640A...1C}, which was calculated using the data from Gaia DR2. This plot shows a negligible (yet positive) shift in the proper motions of objects from Gaia DR2 to Gaia DR3 for OCs. 

\begin{figure}
      \centering
      \subfigure[]{\label{hist_eRV}
      \includegraphics[width=.80\columnwidth]{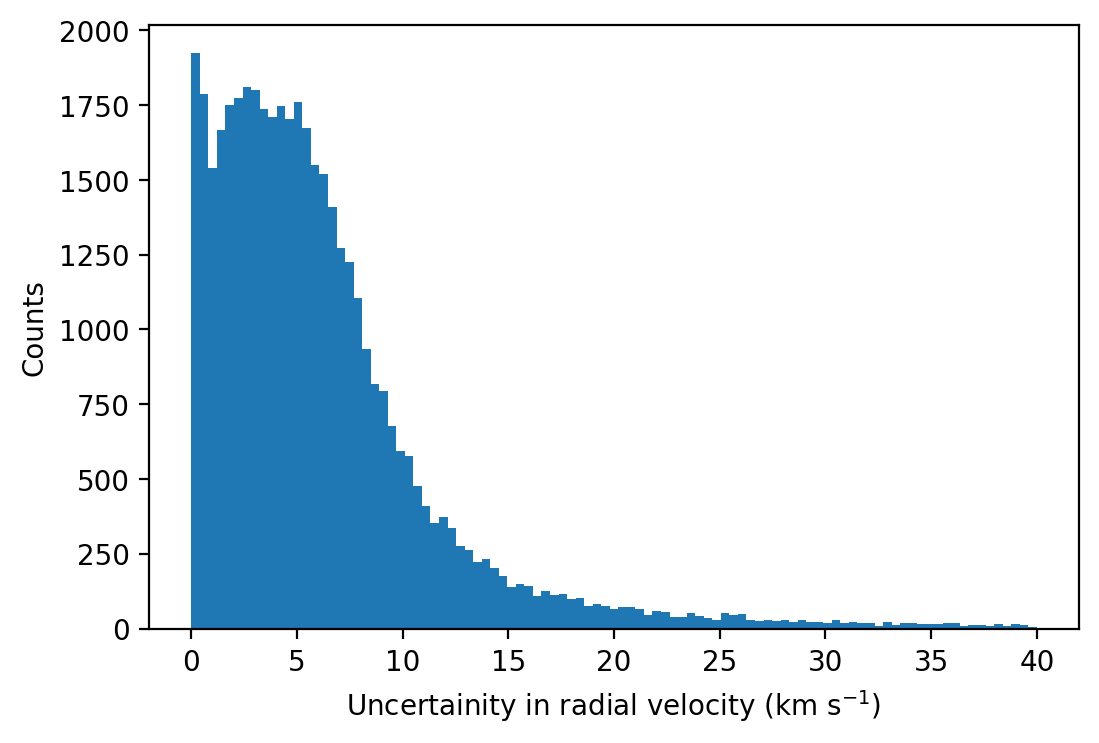}}

     \subfigure[]{\label{hist_RV}
     \includegraphics[width=.80\columnwidth]{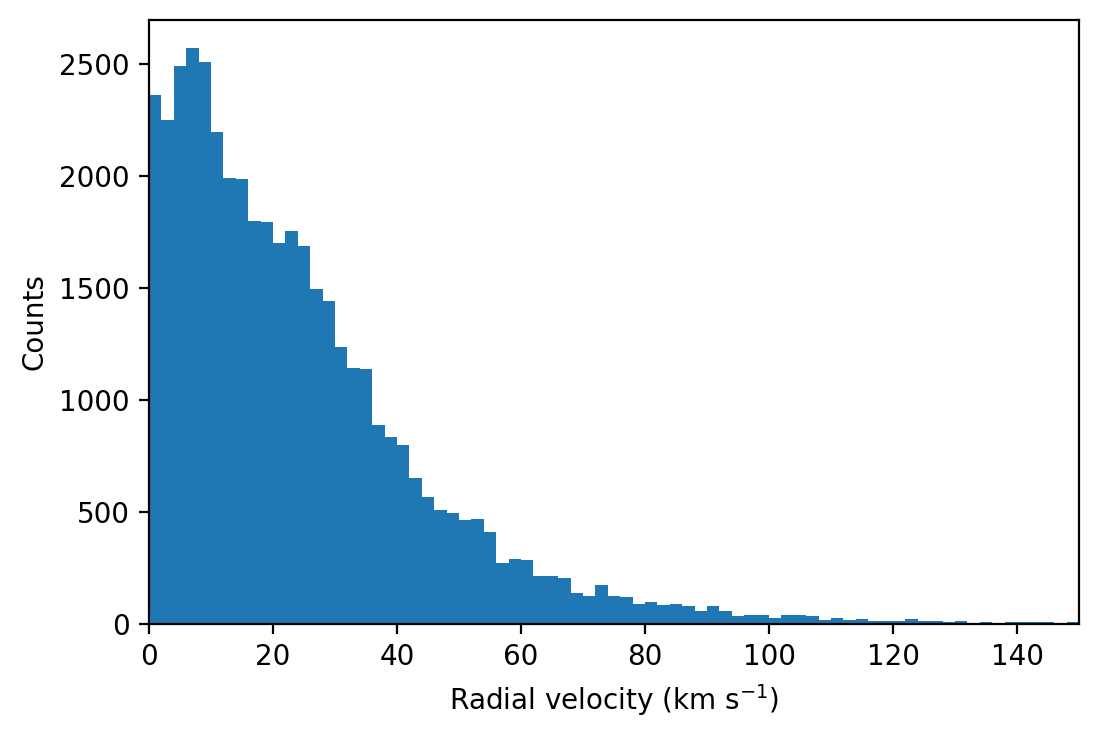}}
     \caption{\ref{hist_eRV}: shows a histogram of uncertainties in radial velocity measurements of 43101 stars. \ref{hist_RV}: shows the histogram of radial velocity measurements.}
\end{figure}
  
We build an ADQL query to extract the nearest Gaia DR3 neighbor for each 
Gaia DR2 source. It was noticed that stars within 1.5 kpc from the Sun had 
multiple matches in DR3 (this could be due to the higher number density in 
this region of the Galaxy). We chose only the star with the least angular 
distance as the cross-match for such cases. After updating all the cluster members for their 
corresponding counterparts in Gaia DR3, parameters for 227,384 stars were 
updated; 43,101 stars (i.e., $\sim19\%$ of the total sample) 
contained radial velocity measurements in Gaia DR3. These 43,101 stars 
belonged to 1,671 OCs. Fig. \ref{hist_eRV} shows the histogram of 
radial velocity uncertainties in these 43,101 stars. The maximum radial 
velocity uncertainty recorded for this sample was 40 km s$^{-1}$, and $93.7\%$ of 
stars had uncertainty less than 15 km s$^{-1}$. 380 stars out of these 43,101 
reported had a radial velocity uncertainty more significant than 20 km s$^{-1}$. 
They had significantly higher effective temperatures (median 8500 K), whereas the rest of the sample with reduced uncertainties had 
comparatively lower effective temperatures (median 5500 K).

From the sample of 43,101 stars belonging to 1,671 OCs, 427 ($0.99\%$) stars had radial velocity more than 102 km s$^{-1}$ and are  3$\sigma$ outliers of the sample (refer Fig. \ref{hist_RV}) distributed in 241 OCs. As these 241 OCs did not homogenously contain higher-velocity stars, the high-velocity stars were not considered for estimating the mean Radial velocity of the OC. 
The cluster radial velocity is calculated by taking a median of the radial velocities of cluster members.
Since the improvement in parallax and magnitude from DR2 to DR3 was not substantial, we adopted the age and distance estimated by \cite{2020A&A...640A...1C} to our sample.
As a result, we have a catalogue of OCs and their kinematical and physical parameters. 
We chose only OCs with at least 10 members with a minimum of 3 stars with recorded radial velocity values. 
Thus, we were left with a sample of 1145 OCs. A short table for this catalogue is given in 
table \ref{tab:catalog}. The whole table will be uploaded electronically. 
In this paper, this catalogue will be termed as "updated or current catalogue". 

\begin{figure*}
\centering
\includegraphics[width=19cm]{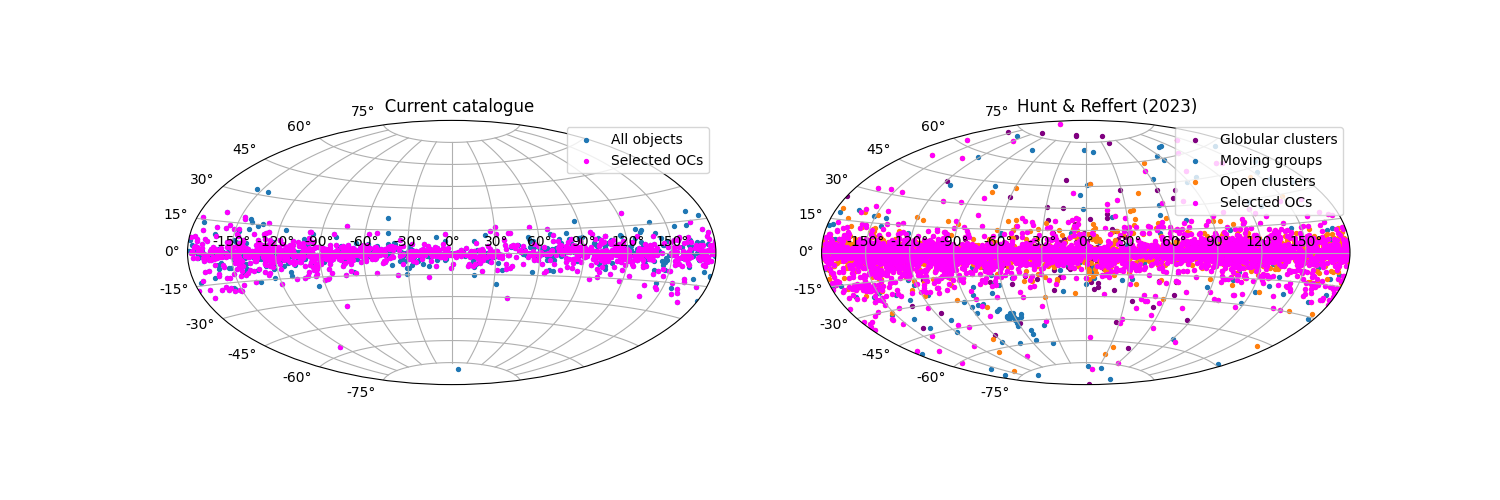}
\caption{Galactic distribution of OCs in the current catalogue (left) and star clusters catalogued
by \citet{2023A&A...673A.114H} (right). We have indicated the selected OCs with magenta points in this figure.
	}
 \label{fig:aitoff}
\end{figure*}

The comprehensive HR23 catalogue has 7167 clusters. Among these, 349 objects are flagged as globular clusters or moving groups. 
We further selected objects with a five-parameter astrometric solution (position, parallax, and proper motion), and the mean radial velocity was calculated using at least three stars. 
After applying these conditions, we were left with a total of 3708 OCs.
We used the distance and age of OCs flagged with a confidence level of 85 in \citet{2023A&A...673A.114H}. 
A short table of HR23, which is similar to \ref{tab:catalog}, is tabulated in \ref{tab:catalog_hunt}.  
The Galactic distributions of the OCs in both catalogues are shown in Fig. \ref{fig:aitoff}. 
This figure shows that the OCs in the updated catalogue are distributed mainly at lower latitudes. In comparison, OCs in HR23 are also present at higher latitudes apart from the dense population near the plane. 
The OCs selected from both catalogues are generally distributed between $-$15 $\deg$ to 15 $\deg$ latitude. We found a total of 816 common OCs between the two catalogues and 329 OCs present in current catalogue are missing from HR23.

\begin{table*}

 \begin{subtable}{}
    \centering
     \caption{The updated catalogue of the OCs using Gaia DR3. RA denotes the right ascension, DEC is 
    declination, pmRA is proper motion in RA, pmDEC is proper motion in DEC, and RV is the radial velocity of
    the OCs. The positions and proper motions are in the ICRS reference system.}
    \begin{tabular}{lccccccc}
    \hline
    \hline
    Name	& Distance & log(age) & RA & DEC & pmRA & pmDEC & RV \\
            & (parsec) &   & (degree) & (degree) & (mas yr$^{-1}$) & (mas yr$^{-1}$) & (km s$^{-1}$) \\
   \hline
Melotte 22 & 128 & 7.89 & 56.6041 & 24.1127 & 19.88 & -45.45 &  5.48  \\
UBC 19 & 416 & 6.84 & 56.3359 & 29.8666 & 2.59 & -5.21 &  17.15   \\
NGC 1333 &  299 & 7.06 & 52.3280 & 31.3175 & 7.06 & -9.82 &  12.50  \\
UBC 199  &  1233 & 9.06 & 67.4256 & 25.4494 & 2.87 & -1.85 & -13.44  \\
Czernik 19   & 2586 & 8.31 & 74.1522 & 28.7838 & 0.80 & -2.11 &  -1.94  \\
Skiff J0507+30.8 &  6088 & 9.39 & 76.7312 & 30.8600 & 0.74 & -0.74 &  -6.44  \\
FSR 0771  &   1529 & 8.45 & 75.8975 & 32.1388 & 1.46 & -4.23 & -13.24  \\
UBC 31 &   341 & 7.42 & 61.3404 & 32.7340 & 3.75 & -5.41 &  18.19  \\
Czernik 18 & 1325 & 8.72 & 66.9343 & 30.9196 & 1.64 & -3.13 & -13.18   \\
FSR 0728 & 1876 & 8.21 & 67.5671 & 38.5244 & 1.80 & -2.30 &  -9.77   \\
COIN-Gaia 11 &  669 & 8.9 & 68.1154 & 39.5161 & 3.46 & -5.69 &  -6.24  \\
COIN-Gaia 10 & 1018 & 7.74 & 68.4593 & 40.8154 & 1.89 & -3.45 &   0.84    \\
COIN-Gaia 20 & 1044 & 7.93 & 78.6551 & 31.6698 & 0.50 & -1.51 &  -0.04  \\
Berkeley 69 & 3410 & 8.9 & 81.1115 & 32.6088 & 0.76 & -1.94 &  39.74   \\
   
    \label{tab:catalog}
    \end{tabular}
    \end{subtable}
    \hspace{0.05\textwidth}
    
    \begin{subtable}{}
    \centering
    \caption{A similar table of parameters for the OCs considered from  \citet{2023A&A...673A.114H}.}
    \begin{tabular}{lccccccc}
    \hline
    \hline
    Name	& Distance & log(age) & RA & DEC & pmRA & pmDEC & RV \\
            & (parsec) &   & (degree) & (degree) & (mas yr$^{-1}$) & (mas yr$^{-1}$) & (km s$^{-1}$) \\
   \hline
   Melotte 22 & 134.859 & 8.29 & 56.6798 & 24.1085 & 19.96 & -45.46 & 05.34 \\
   UBC 19 & 393.509 &  6.59 & 56.3306 & 29.7416 & 02.67 & -05.24 & 18.20 \\
   NGC 1333 & 291.099 & 6.95 & 52.3282 & 31.3582 & 06.84 & -09.71 & 21.05 \\
   UBC 199  & 1143.070 & 8.95 & 67.4366 & 25.4189 & 02.91 & -01.84 & -11.58 \\
   FSR 0771 & 1554.363 & 8.49 & 75.9211 & 32.1453 & 01.50 & -04.26 & -06.51 \\
   UBC 31 & 368.515 & 7.05 & 60.5210 & 32.1781 & 03.63 & -05.26 & 24.98 \\
   FSR 0728  & 1737.361 & 8.76 & 67.4620 & 38.4982 & 01.92 & -02.27 & 24.89 \\
   COIN-Gaia 11  &  643.450 & 8.72 & 68.1497 & 39.5332 & 03.44 & -05.75 & -04.94 \\
   COIN-Gaia 10 & 1010.894 & 8.61 & 68.4584 & 40.4996 & 01.97 & -03.45 & 08.09 \\
   COIN-Gaia 20 & 1031.823 & 8.06 & 78.6360 & 31.7201 & 00.54 & -01.50 & -00.45 \\
   Berkeley 69  & 3212.265 & 8.81 & 81.0909 & 32.6082 & 00.74 & -01.95 & 31.68 \\
   Berkeley 70  & 4049.934 & 9.16 & 81.4544 & 41.9501 & 00.83 & -01.87 & -10.82 \\
   Berkeley 71  & 3561.109 & 8.78 & 85.2418 & 32.2727 & 00.63 & -01.66 & -08.45 \\
   Berkeley 76  & 5325.960 & 9.19 & 106.6595 & -11.7236 & -00.58 & 01.42 & 108.11 \\
   Berkeley 78  & 4419.755 & 9.55 & 110.9088 & 05.3713 & -00.06 & -01.20 & 71.54 \\
    \label{tab:catalog_hunt}
    \end{tabular}
    \end{subtable}
   
\end{table*}

\section{Orbits of open clusters}\label{sec:orbits}

We used the updated/current catalogue and OCs selected from HR23 to study the distribution and motion of OCs in the Galactic disk. 
For this, we first transformed the heliocentric position and velocity components of all the OCs into their Galactocentric counterparts and then 
traced their trajectory in the Galaxy. For this purpose, we used the Python-based module \textbf{galpy} \citep{2015ApJS..216...29B}. 
It uses Astropy units and performs coordinates transformation internally. 
\textit{galpy} contains a general framework depicting various Galactic potentials that can be used as isolated or a mixture of multiple potentials.
We used the model \textbf{MWPotential2014} to estimate the potentials of the Galaxy best. 
This model is a superposition of power-law spherical potential with an exponential cutoff \citep{cardone2005spherical} 
representing the Galactic Bulge, Miyamoto Nagai potential \citep{1975PASJ...27..533M} representing the Galactic Disk, and 
the NFW potential \citep{1997ApJ...490..493N} representing the Galactic Halo. 
The parameters of these three potentials have been obtained after fitting dynamical data from \cite{holmberg2000local}, \cite{xue2008milky}, \cite{bovy2012local} and \cite{Zhang_2013} to best portray the realistic description of the Milky Way on small and large scales.
For this analysis, we assumed the location of the Sun at a distance of 8.178 kpc from the Galactic center with a velocity of 220 km s$^{-1}$ in the direction of Galactic rotation. 

\begin{figure*}
    \centering
    \subfigure[]{
    \includegraphics[width=.48\columnwidth]{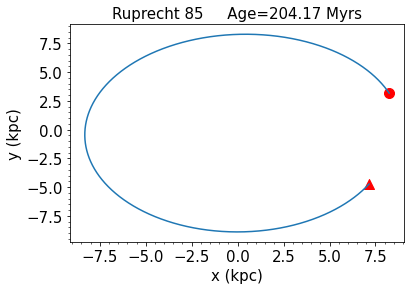}}
    \subfigure[]{
    \includegraphics[width=.48\columnwidth]{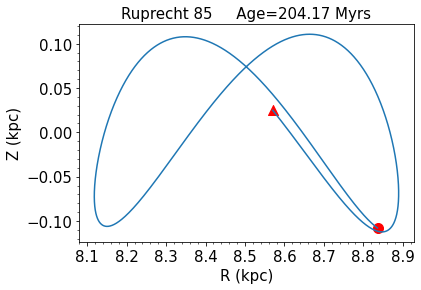}}
    \subfigure[]{
    \includegraphics[width=.48\columnwidth]{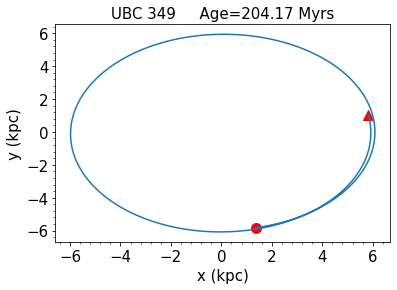}}
    \subfigure[]{
    \includegraphics[width=.48\columnwidth]{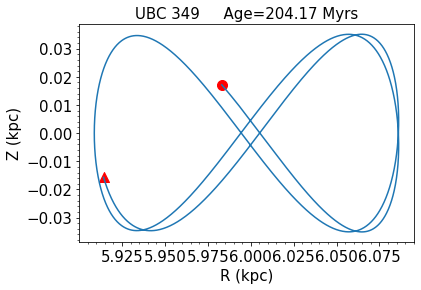}}
    \subfigure[]{
    \includegraphics[width=.48\columnwidth]{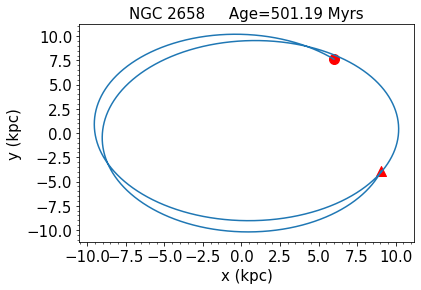}}
    \subfigure[]{
    \includegraphics[width=.48\columnwidth]{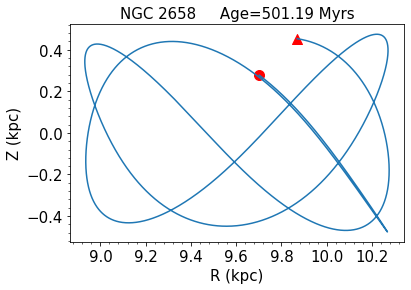}}
    \subfigure[]{
    \includegraphics[width=.48\columnwidth]{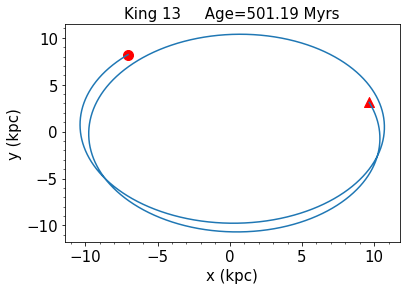}}
    \subfigure[]{
    \includegraphics[width=.48\columnwidth]{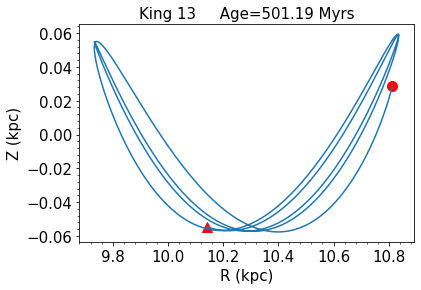}}
    \subfigure[]{
    \includegraphics[width=.48\columnwidth]{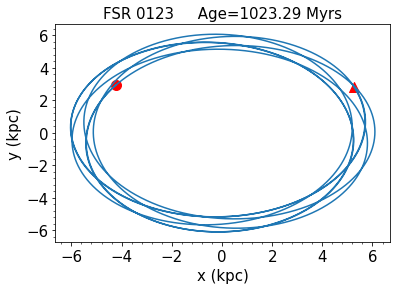}}
    \subfigure[]{
    \includegraphics[width=.48\columnwidth]{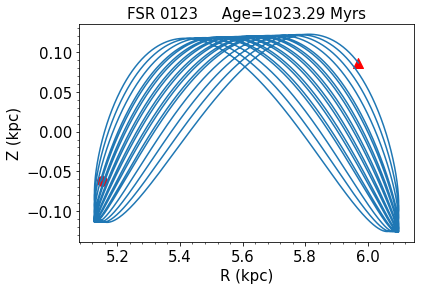}}
    \subfigure[]{
    \includegraphics[width=.48\columnwidth]{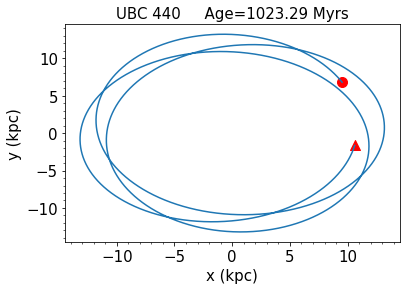}}
    \subfigure[]{
    \includegraphics[width=.48\columnwidth]{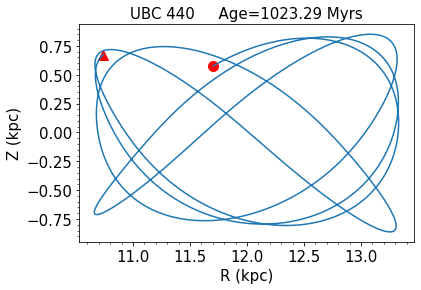}}
    \subfigure[]{
    \includegraphics[width=.48\columnwidth]{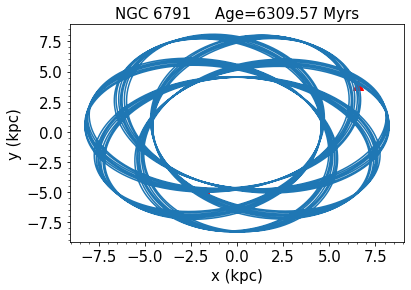}}
    \subfigure[]{
    \includegraphics[width=.48\columnwidth]{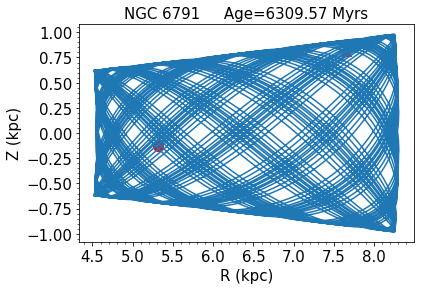}}
    \subfigure[]{
    \includegraphics[width=.48\columnwidth]{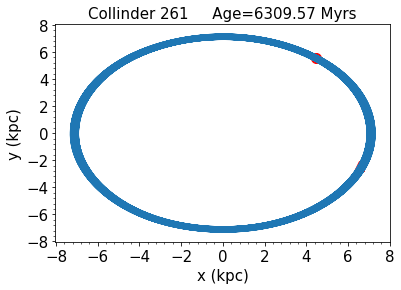}}
    \subfigure[]{
    \includegraphics[width=.48\columnwidth]{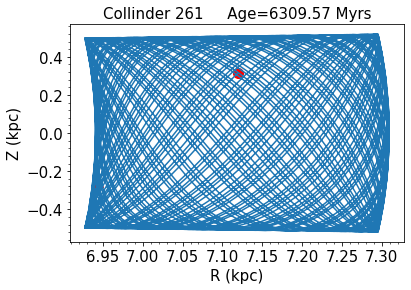}}
    \qquad
    \caption{Above are the orbits of some OCs from different age groups. Their present-day and birth positions are marked by a red triangle and a red dot, respectively. Rows show orbits for OCs nearly 200 Myr, 500 Myr, 1Gyr, and then OCs older than 6 Gyr old.}
    \label{many orbits}
\end{figure*}

We back-integrated the cluster orbits for our sample using the parameters determined above to a period equal to the 
respective cluster age at a time step of every 0.1 Myr. The orbits of eight OCs in $\sim$ 200 Myr - 6 Gyr age range are displayed in Fig.
\ref{many orbits}. 
The birth and present-day positions of these OCs are denoted by red dot and triangle, respectively.
The Galactocentric position, velocity components, and the orbital parameters of 15 OCs mentioned in table \ref{tab:catalog}, from the updated catalogue, are listed in \ref{tab:orbit}. 
The complete table for 1145 OCs will be provided in electronic form.

We also calculated the orbits of 3708 OCs selected from HR23 using the same method, and a short table of the output parameters is tabulated in \ref{tab:orbit_hunt} (a complete table will be made available in the online form). 
We found that cluster Palomar 12 is flagged as an open cluster in HR23, which is a globular cluster and traces a maximum height from the Galactic mid-plane ($Z_{max}$) $>$ 30 kpc. 
We excluded Palomar 12 and 30 OCs having radial distance from the Galactic center ($R$)  $<$ 5 kpc to avoid the inclusion of False OCs.
The remaining 3677 OCs are used for further analysis. The common OCs between the final set catalogues are same as earlier as there are no OCs present in current catalogue having $R < 5$ kpc and $Z_{max} > 30$ kpc so the final catalogue of orbital parameters include a total of 4006 OCs. As there are differences in the parameter values, the two catalogues are not merged. We proceed with analysing both catalogues and comparing the results, as given below. 

\begin{table*}
\hspace{0cm}
%%\begin{subtable}{}
\begin{tiny}
\centering
\begin{tabular}{lrrrrrrrrrrrrrrr}
\hline
\hline
Name         & e & $R_{A}$ & $R_{P}$ & $Z_{max}$ & $L_{Z}$ & $V_{X}$   &  $V_{Y}$   & $V_{Z}$ &    X   &  Y   &    Z     &$X_{birth}$&$Y_{birth}$& $Z_{birth}$& R  \\
 &   & (kpc) & (kpc) &  (kpc) & (km s$^{-1}$ kpc)  &   (km s$^{-1}$) &  (km s$^{-1}$) & (km s$^{-1}$) &  (kpc) & (kpc) & (kpc) & (kpc) & (kpc) & (kpc) & (kpc)  \\
\hline
Melotte 22              & 0.07 &  8.46 &  7.35 & 0.09 & 1735.96 &  -4.56    &   205.57  & -6.23  &  8.44 & 0.03 & -0.03 & -4.09 & -6.11 & -0.06 &  8.44  \\
UBC 19                  & 0.00 &  8.71 &  8.65 & 0.12 & 1972.17 &   6.12    &   226.67  & -2.66  &  8.70 & 0.12 & -0.12 &  8.52 & -1.48 & -0.09 &  8.71  \\
NGC 1333                & 0.00 &  8.59 &  8.52 & 0.08 & 1895.32 &   4.20    &   220.68  & -1.82  &  8.59 & 0.10 & -0.08 &  8.16 & -2.45 & -0.04 &  8.59 \\
UBC 199                 & 0.06 &  9.94 &  8.78 & 0.46 & 2016.91 & -20.07    &   211.80  & 16.10  &  9.51 & 0.15 & -0.31 & -3.35 & -8.15 &  0.11 &  9.51 \\
Czernik 19              & 0.05 & 10.95 & 10.01 & 0.37 & 2238.05 & -11.49    &   205.58  & -0.50  & 10.87 & 0.27 & -0.37 & -4.62 &  9.65 &  0.36 & 10.88  \\
Skiff J0507+30.8        & 0.03 & 14.49 & 13.78 & 0.78 & 2908.56 & -13.77    &   202.04  & 12.31  & 14.35 & 0.66 & -0.59 &  5.87 & 12.55 & -0.66 & 14.37  \\
FSR 0771                & 0.10 & 10.03 &  8.20 & 0.13 & 1969.73 & -20.97    &   199.78  & -1.52  &  9.84 & 0.21 & -0.13 &  8.13 & -2.95 & -0.06 &  9.84   \\      
UBC 31                  & 0.01 &  8.65 &  8.53 & 0.06 & 1962.86 &   8.25    &   227.09  &  0.65  &  8.65 & 0.09 & -0.06 &  6.48 & -5.56 &  0.01 &  8.65  \\      
Czernik 18              & 0.06 &  9.86 &  8.67 & 0.28 & 2005.25 & -20.52    &   208.36  &  4.49  &  9.60 & 0.26 & -0.26 &  8.63 &  0.83 & -0.04 &  9.60  \\
FSR 0728                & 0.05 & 10.13 &  9.10 & 0.23 & 2075.40 & -13.03    &   204.55  &  6.16  & 10.11 & 0.55 & -0.20 & -7.98 &  5.51 &  0.18 & 10.13  \\
COIN-Gaia 11            & 0.05 &  9.02 &  8.25 & 0.07 & 1889.41 & -11.19    &   210.51  &  3.72  &  8.97 & 0.20 & -0.05 & -0.58 & -8.51 &  0.05 &  8.97  \\
COIN-Gaia 10            & 0.01 &  9.30 &  9.08 & 0.07 & 1998.47 &  -4.87    &   214.87  &  2.67  &  9.29 & 0.32 & -0.06 &  2.47 & -8.74 &  0.07 &  9.30  \\
COIN-Gaia 20            & 0.04 & 10.23 &  9.37 & 0.10 & 2106.15 & -10.52    &   224.75  &  4.95  &  9.37 & 0.11 & -0.05 & -3.08 & -9.76 &  0.07 &  9.37  \\
Berkeley 69             & 0.14 & 12.93 &  9.76 & 0.09 & 2372.75 &  31.40    &   203.29  & -1.35  & 11.72 & 0.33 & -0.08 & -8.89 & -4.46 & -0.00 & 11.73  \\
             
\label{tab:orbit}
\end{tabular}
\end{tiny}
\caption{The orbital parameters calculated from the orbit integration and the position and velocity coordinates of the OCs in the updated catalogue, where $e$ is the eccentricity of the orbit, $R_{A}$ and $R_{P}$ are apogalactic and perigalactic radii, $Z_{max}$ is the maximum height attained by the cluster from the Galactic disk in its total travel so far, $L_{Z}$ is the angular momentum, ($V_{X}, V_{Y}, V_{Z}$) and ($X, Y, Z$) are the velocity and position cartesian coordinates of the OCs, ($X_{birth}, Y_{birth}, Z_{birh}$) are the position of the OCs at the time of their birth in the Galaxy and $R$ is the radial distance of the cluster from the Galactic center.}
%%\end{subtable}
\hspace{0.05\textwidth}
\end{table*}

\begin{table*}
\hspace{0cm}
%%\begin{subtable}{}
\begin{tiny}
%%\centering
\begin{tabular}{lrrrrrrrrrrrrrrr}
\hline
\hline
 Name         & e & $R_{A}$ & $R_{P}$ & $Z_{max}$ & $L_{Z}$ & $V_{X}$   &  $V_{Y}$   & $V_{Z}$ &    X   &  Y   &    Z     &$X_{birth}$&$Y_{birth}$& $Z_{birth}$& R  \\
 &   & (kpc) & (kpc) &  (kpc) & (km s$^{-1}$ kpc)  &   (km s$^{-1}$) &  (km s$^{-1}$) & (km s$^{-1}$) &  (kpc) & (kpc) & (kpc) & (kpc) & (kpc) & (kpc) & (kpc)  \\
\hline
Melotte22  &  0.08 &  8.31 & 7.08 & 0.10 & 1692.93 & -4.58 & 203.99 & -6.73 & 8.30 & 0.03 & -0.03 & 6.31 & 4.74 & 0.06 & 8.30  \\
UBC 19     &  0.00 &   8.53 & 8.50 & 0.11 & 1939.13 & 7.02 & 227.38 & -2.74 & 8.53 & 0.11 & -0.11 & 8.46 & -0.80 & -0.10 & 8.53 \\
NGC 1333   &  0.01 &   8.43 & 8.31& 0.08 & 1890.59 & 11.37 & 224.36 & -4.71 & 8.43 & 0.10 & -0.08 & 8.09 & -1.94 & -0.02 & 8.43 \\
UBC 199    &  0.06 & 9.69 & 8.66 & 0.42 & 1979.80 & -18.53 & 213.28 & 15.45 & 9.27 & 0.14 & -0.29 & -7.06 & -5.40 & 0.34 & 9.27 \\
FSR 0771   &  0.09 &   9.78 & 8.19 & 0.13 & 1943.84 & -14.23 & 199.86 & -2.19 & 9.71 & 0.21 & -0.13 & 1.90 & -8.01 & -0.02 & 9.71 \\
UBC 31 & 0.01 &   8.52 & 8.36 & 0.08 & 1945.10 & 14.52 & 228.50 & -1.60 & 8.52 & 0.10 & -0.08 & 7.97 & -2.51 & -0.04 & 8.52 \\
FSR 0728 & 0.11 &   10.97 & 8.85 & 0.21 & 2113.40 & 19.58 & 216.06 & 3.14 & 9.83 & 0.51 & -0.18 & 10.56 & -2.85 & -0.21 & 9.84 \\
COIN-Gaia 11 & 0.04 &   8.84& 8.18 & 0.07 & 1861.34 & -10.17 & 211.56 & 3.60 & 8.79 & 0.19 & -0.04 & 2.22 & -8.45 & -0.02 & 8.79 \\
COIN-Gaia 20  & 0.04 &   10.06& 9.20 & 0.10 & 2068.99 & -10.90 & 224.72 & 5.17 & 9.20 & 0.11 & -0.05 & -8.17 & -5.53 & -0.08 & 9.20 \\
 Berkeley 69  &  0.11 &   12.22& 9.79 & 0.08 & 2318.71 & 23.19 & 204.50 & -1.10 & 11.37 & 0.31 & -0.07 & 11.30 & -4.08 & 0.02 & 11.38 \\
Berkeley 70  &  0.10 & 12.16& 10.04 & 0.29 & 2338.46 & -12.36 & 192.11 & -0.22 & 12.11 & 0.92 & 0.29 & -12.10 & 0.97 & -0.28 & 12.15 \\
Berkeley 71  & 0.07 &   12.01& 10.50 & 0.09 & 2378.78 & -17.71 & 202.44 & 1.33 & 11.73 & 0.21 & 0.09 & 3.95 & 10.38 & -0.05 & 11.73 \\
Berkeley 78  & 0.16 & 12.14 & 8.76 & 0.81 & 2174.14 & 39.13 & 175.44 & 7.05 & 11.88 & -2.30 & 0.77 & -2.75 & 8.55 & 0.61 & 12.10\\
\end{tabular}
\end{tiny}
\caption{A similar table of the orbital parameters for the OCs taken from HR23.}
\label{tab:orbit_hunt}
%%\end{subtable}
\end{table*}

\begin{figure*}
\centering
\subfigure[]{\label{hist_c}
     \includegraphics[width=8.5cm]{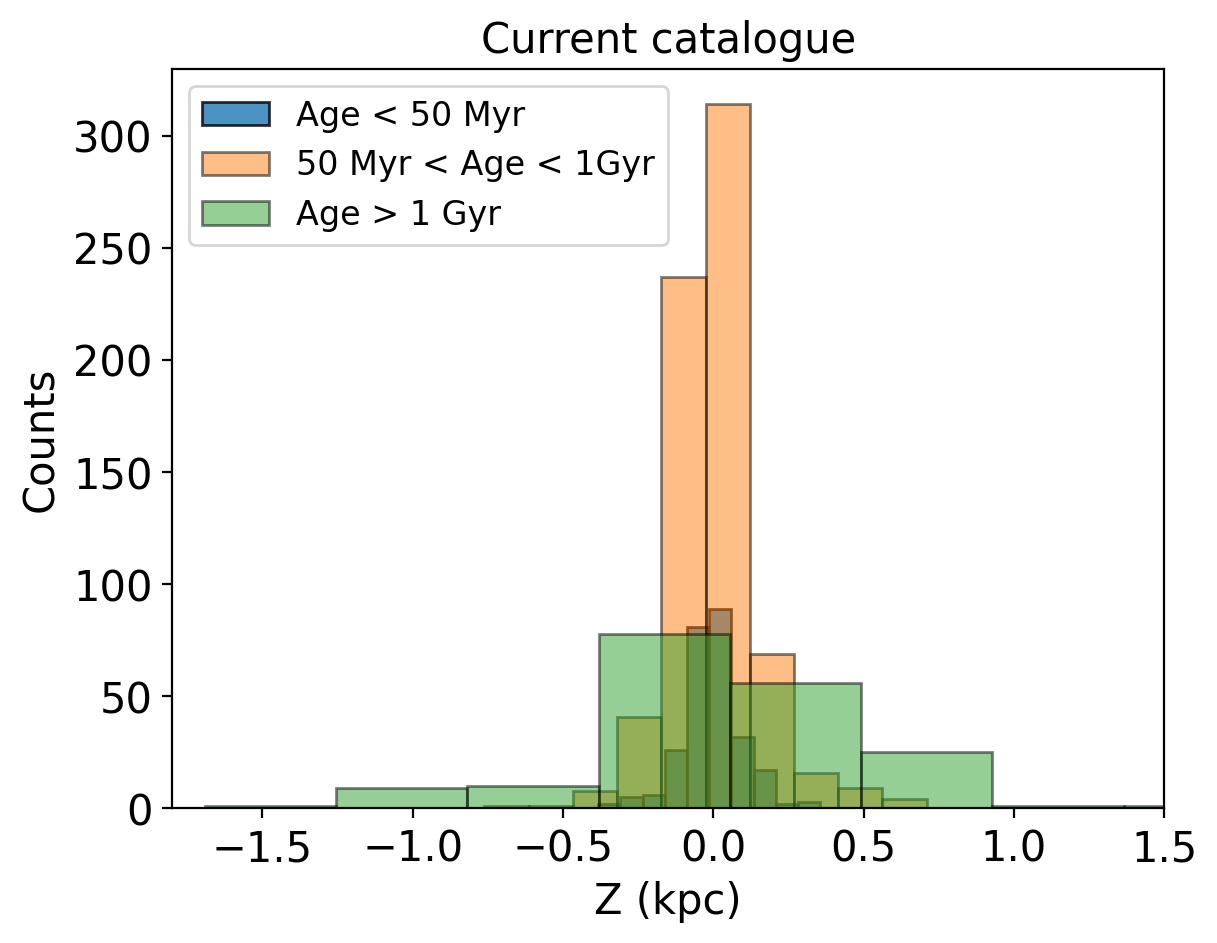}
     \includegraphics[width=8.5cm]{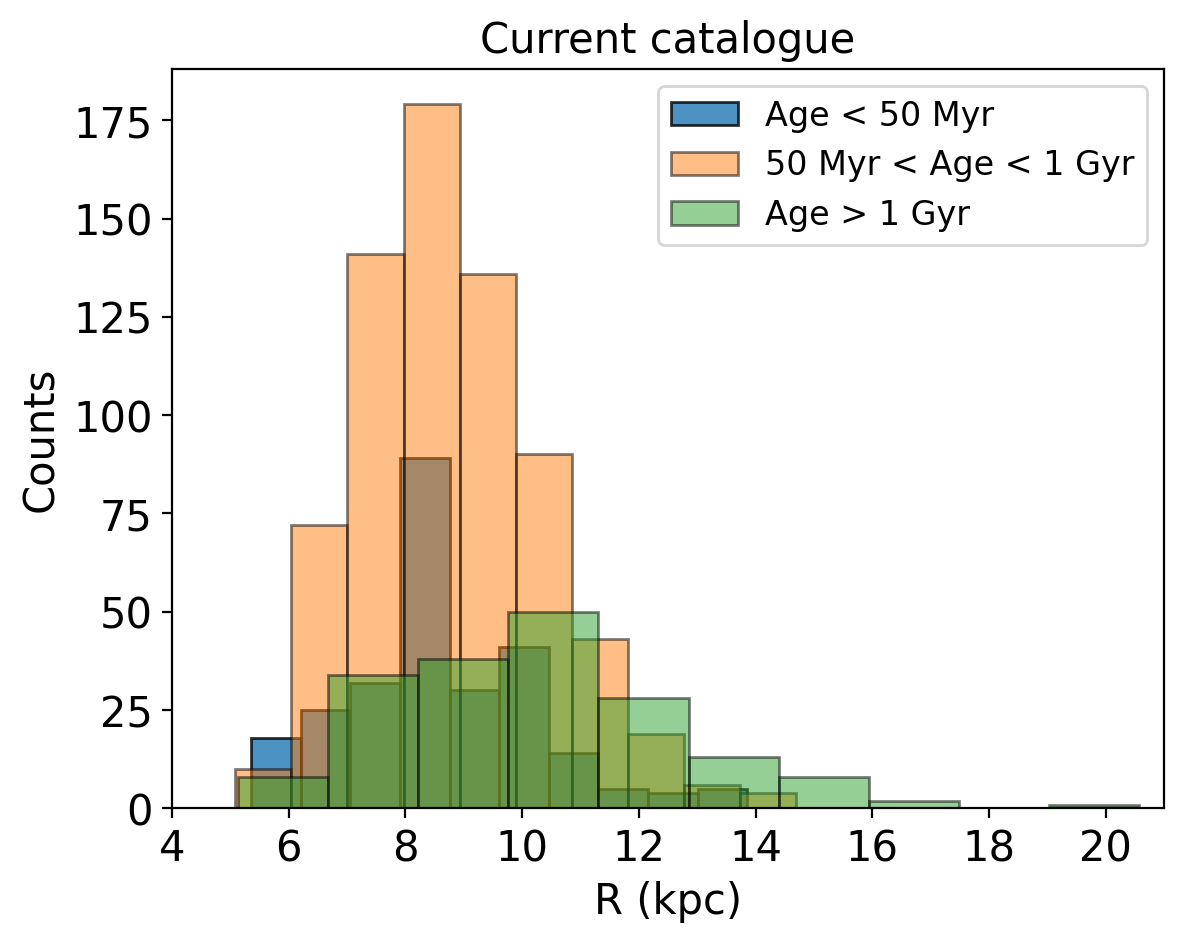}}

\subfigure[]{\label{hist_hunt}
    \includegraphics[width=8.5cm]{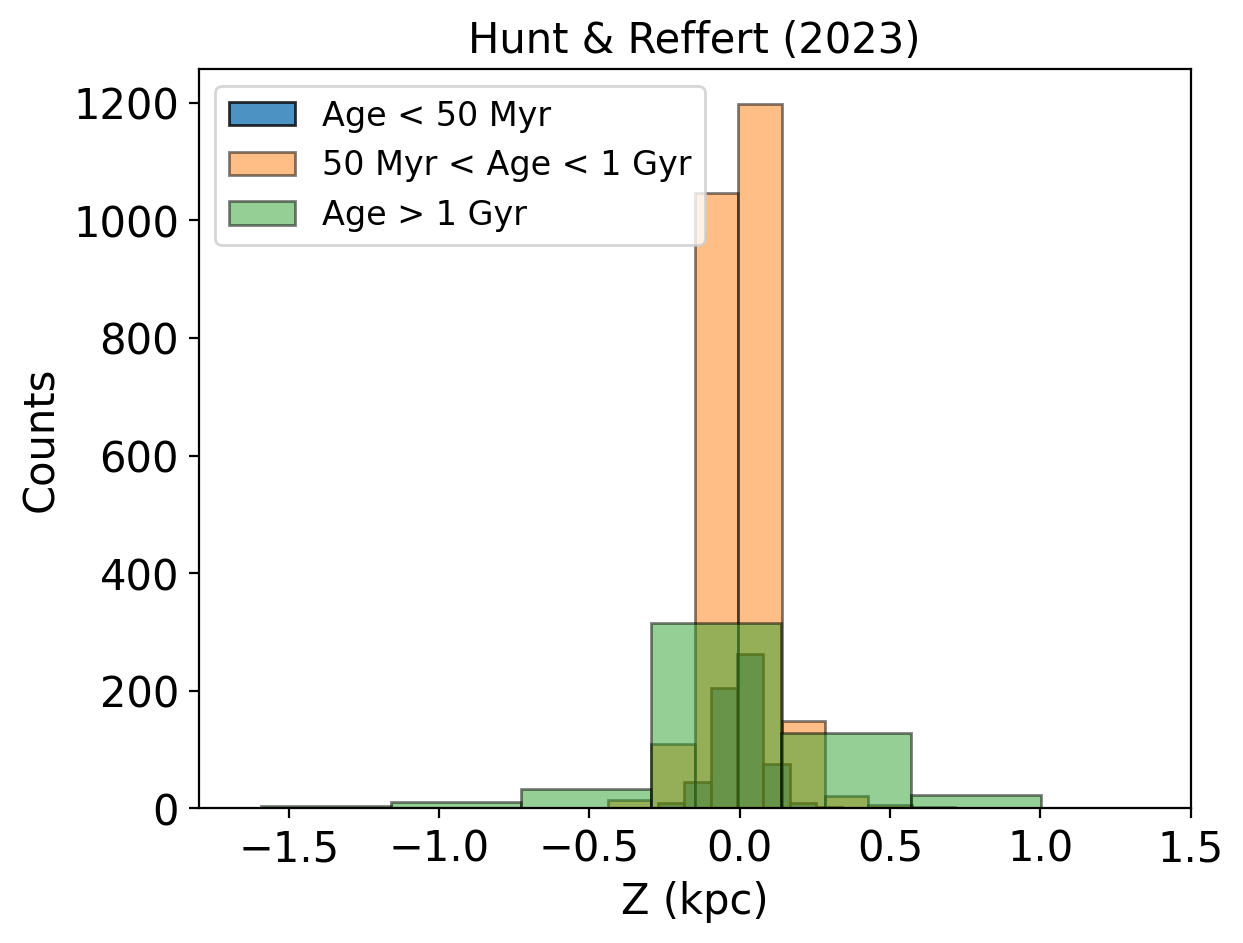}
    \includegraphics[width=8.5cm]{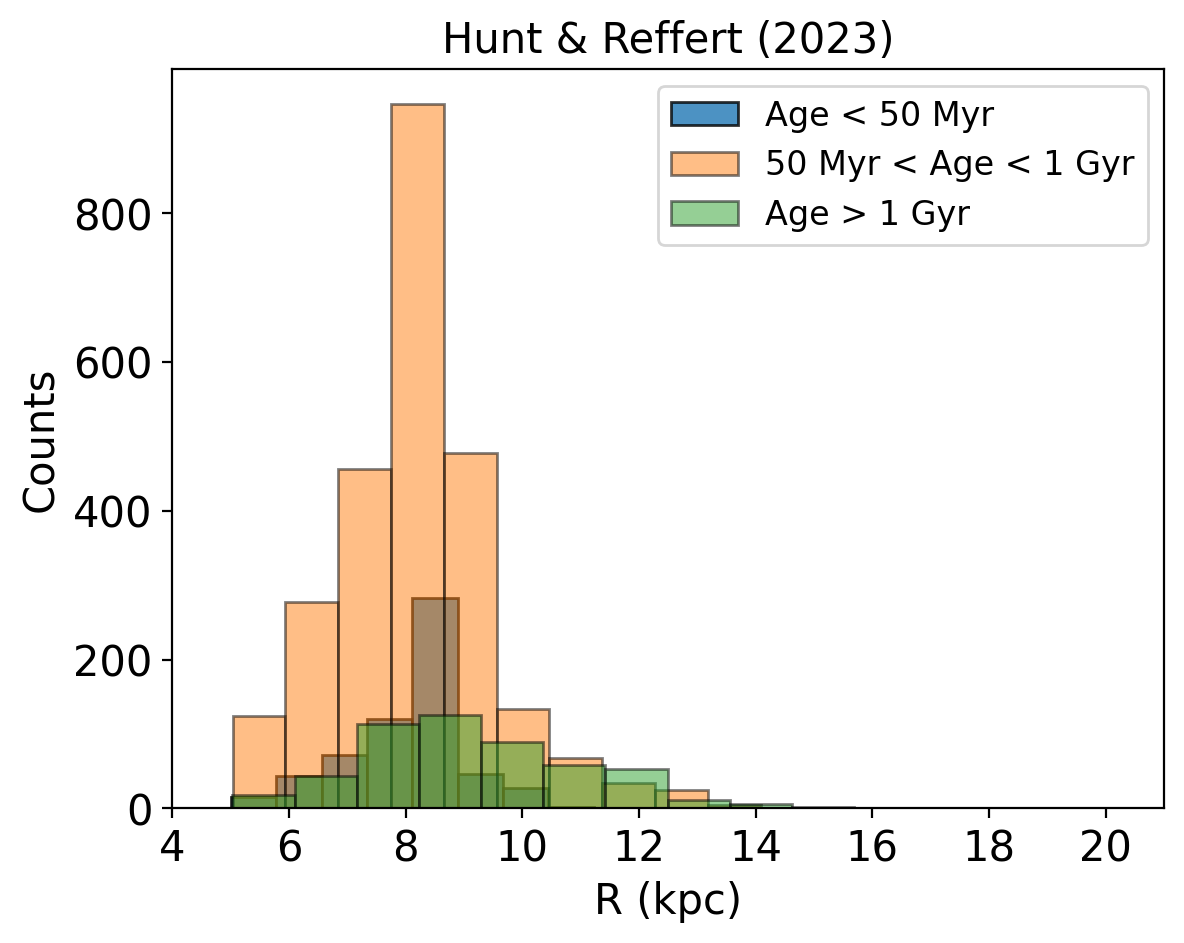}}
    \qquad
\caption{ The left panel of \ref{hist_c} shows a histogram of vertical height from the Galactic mid-plane, whereas the right panel shows our sample's histogram of Galactocentric radius for OCs in current catalogue. The left and right panels of \ref{hist_hunt} show the similar figure for HR23.} 
\label{hist}
\end{figure*}

\section{Results}\label{sec:results}

Orbits and corresponding orbital parameters such as eccentricity, perigalactic and apogalactic distance, and Z$_{max}$ while orbiting the Galactic center were derived for 1145 and %3708 
3677 OCs in both catalogues using the above-discussed method. 
To interpret the results, we divided the sample OCs into three groups: young OCs (younger than 50 Myr), intermediate age OCs (aged 
from 50 Myr to 1 Gyr), and old OCs (older than 1 Gyr). 
These three groups consist of 263 (611), 700 (2547), and 182 (519) OCs, respectively, in the updated (HR23) catalogue respectively.

Fig. \ref{hist_c} shows a histogram of the present-day vertical heights and $R$ of OCs for the current catalogue, and Fig. \ref{hist_hunt} shows similar histograms for HR23. 
These figures give us an idea of the present-day distribution of  the sample OCs in the Galaxy and show that most young and intermediate-age OCs are located close to the Sun. 
In contrast, the older OCs are distributed radially from the inner region to the disk's outer region. 
Similarly, young and intermediate age OCs are predominantly found closer to the Galactic mid-plane. 
In contrast, older OCs are distributed with more considerable vertical heights with respect to the Galactic mid-plane. 

\begin{figure*}[htp]
    \centering
    \includegraphics[width=.9\columnwidth]{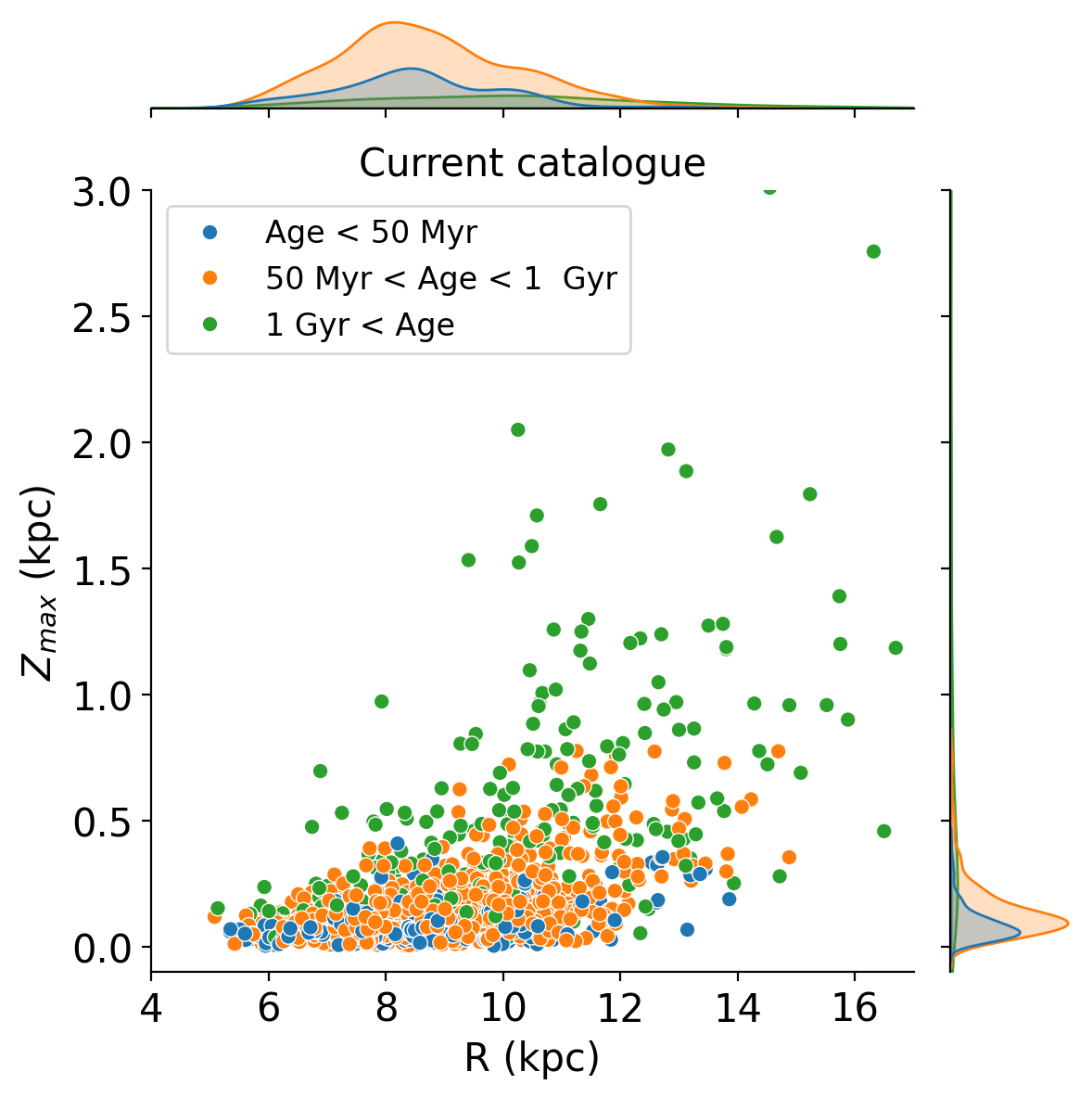}
    \includegraphics[width=.9\columnwidth]{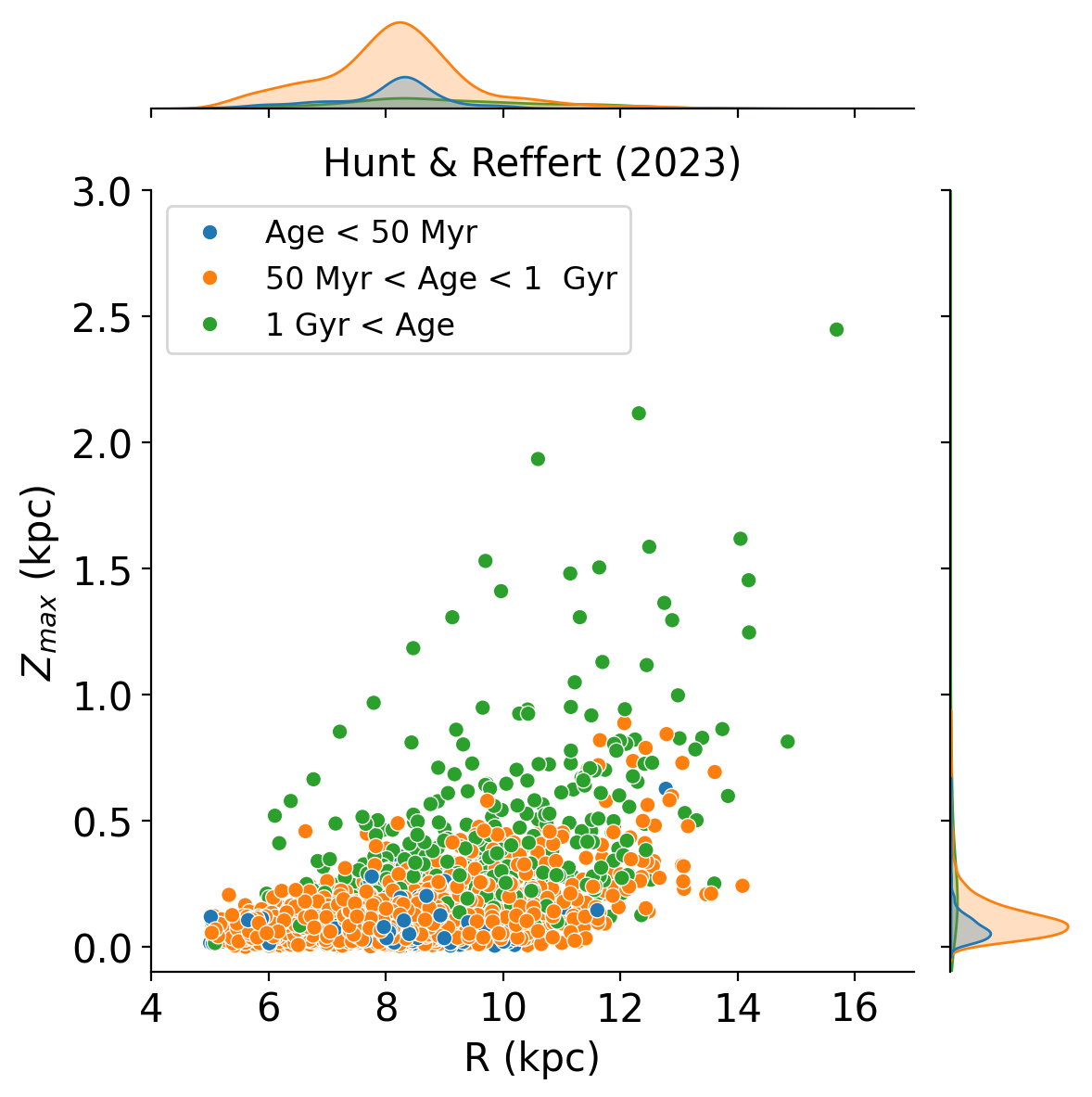}
    \caption{A scatter plot between the maximum height ($Z_{max}$) reached by OCs in its orbit and the Galactocentric radius ($R$) for the three groups of OCs, as stated in the legend for the sources from the two catalogues.}
    \label{scatter_z}
\end{figure*}

The Z$_{max}$ values of OCs can determine the maximum vertical height of the Galactic disk as a function of age and $R$. 
Hence, we plotted the Z$_{max}$ of all the selected OCs in both catalogues as a function of their $R$ in Fig. \ref{scatter_z}. 
We found a distinct pattern between the maximum vertical height and the $R$ for the three age groups. 
Only the older OCs are able to reach larger vertical heights from the disk, while the young OCs are confined to the Galactic mid-plane. 
Also, as we proceed farther from the galactic center in a radial direction, the capability of any cluster to move in an orbit with a larger vertical height increases.
The overall trend of disk thickening is the same for both OC catalogues.

\begin{figure}
    \centering
    \subfigure[]{\label{kde r}
    \includegraphics[width=.8\columnwidth]{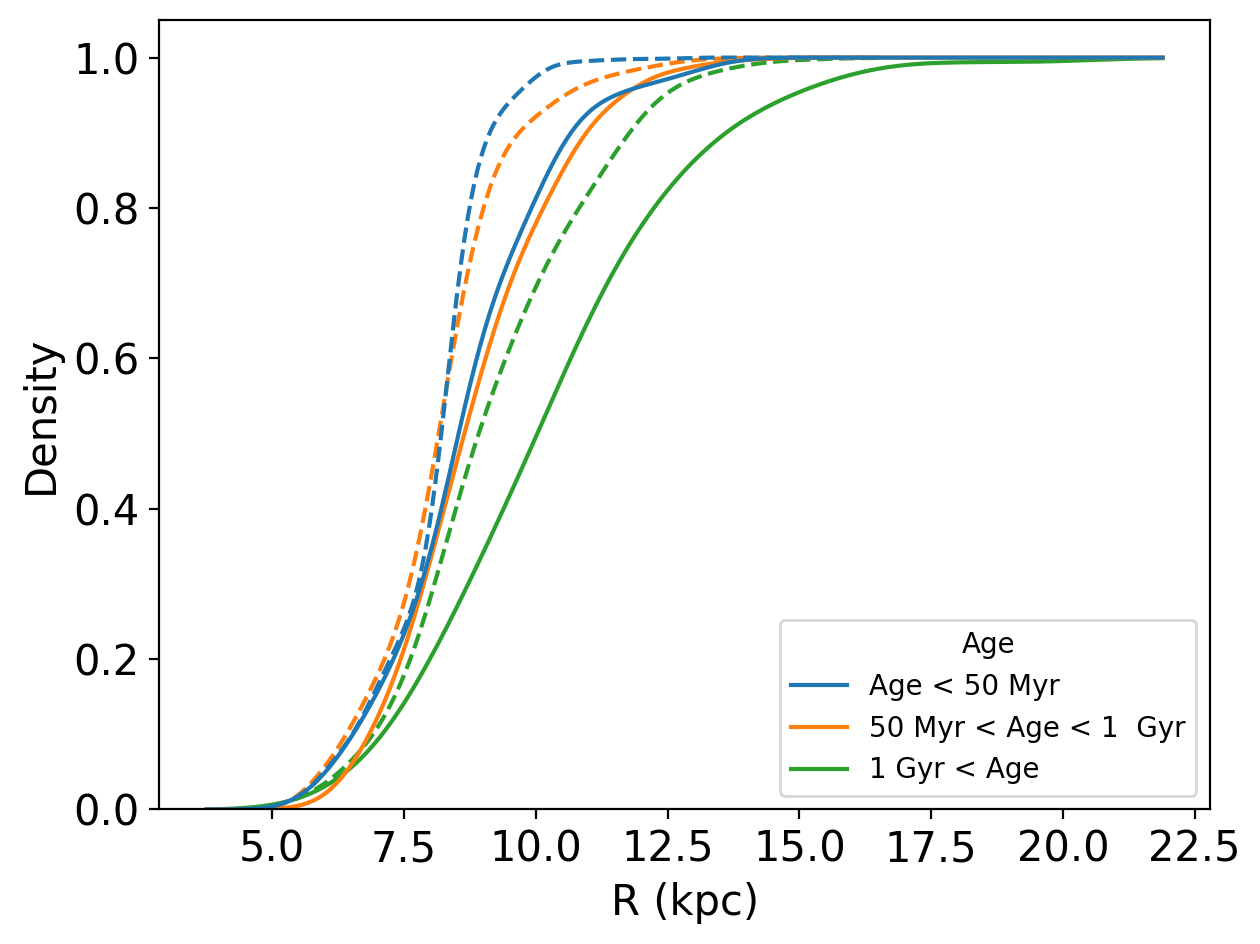}}
    \subfigure[]{\label{kde zmax}
    \includegraphics[width=.8\columnwidth]{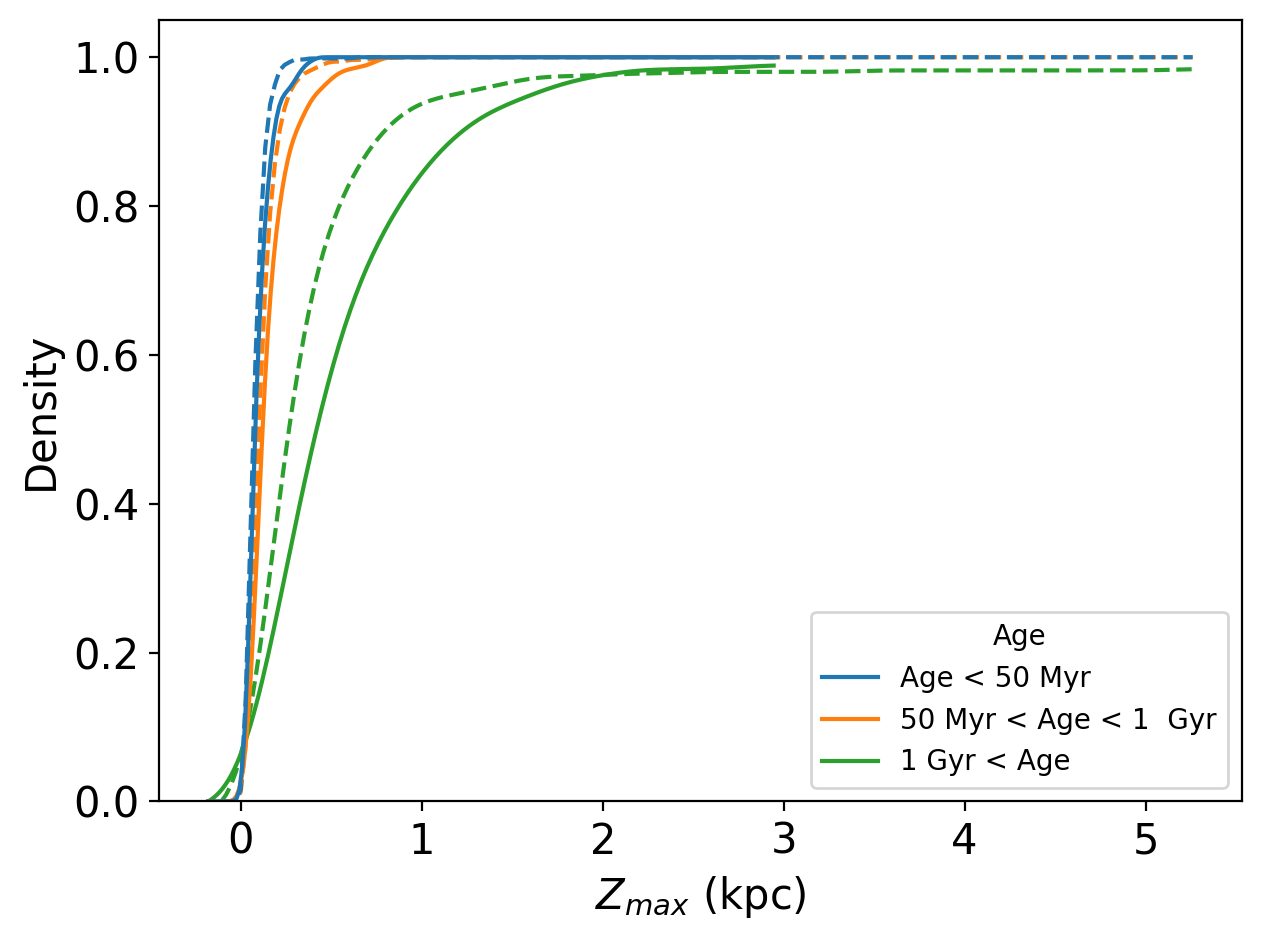}}
    \caption{The cumulative kernel density distribution for 1145 OCs across Galactocentric radius $R$ (panel \ref{kde r}) and maximum vertical height from the Galactic mid-plane $Z_{max}$ (panel \ref{kde zmax}) for different age bins. The Solid lines represent current catalogue while the dotted lines represent catalogue from \citet{2023A&A...673A.114H}.}
     \label{kde}
\end{figure}

To explore the relative distribution of OCs in various age groups, we plotted the cumulative kernel density plots for $R$ and $Z_{max}$ as shown in Fig. \ref{kde}.
The solid lines in these figures represent the OCs in the current catalogue, while the dotted lines represent OCs in HR23.
The figure in the upper panel shows that for OCs younger than 1 Gyr, the distribution across radial distance appears to be similar. 
For OCs older than 1 Gyr, we note an extended distribution, with more older OCs in the outer part of the disk. The OCs in both catalogues show a similar trend, but the OCs in the HR23 catalogue are radially located relatively inside.
In the lower panel of the figure, it can be noted that OCs younger than 1 Gyr have a very similar distribution of vertical height. 
In contrast, OCs older than 1 Gyr can reach significantly higher vertical heights, as seen from both catalogues. 
The figure suggests that the maximum vertical height that can be reached by OCs younger than 1 Gyr is $\sim$ 0.5 kpc, whereas OCs older than 1 Gyr can achieve up to a maximum vertical height of $\sim$ 2 kpc. 
The OCs in both catalogues follow a similar trend; the only difference is that HR23 has about three times the number of OCs in all three age ranges.

\begin{figure*}
    \centering
    \includegraphics[width=8cm, height=6.2cm]{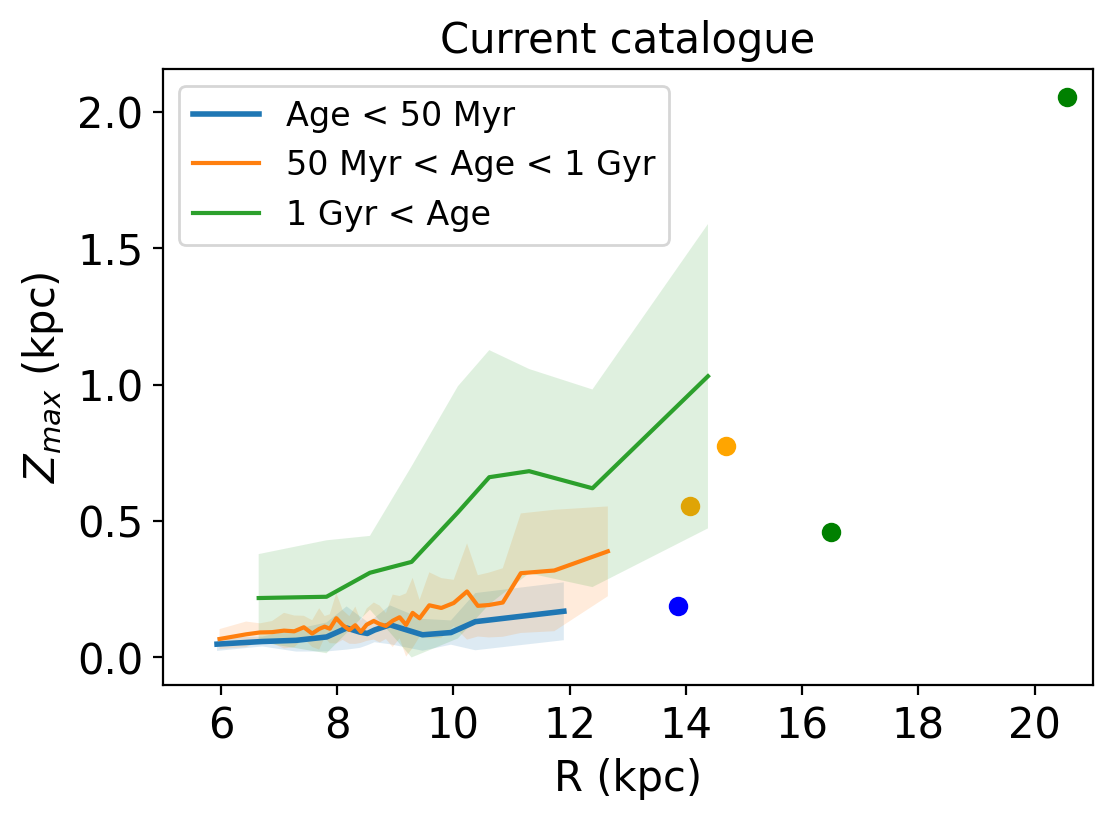}
    \includegraphics[width=8cm]{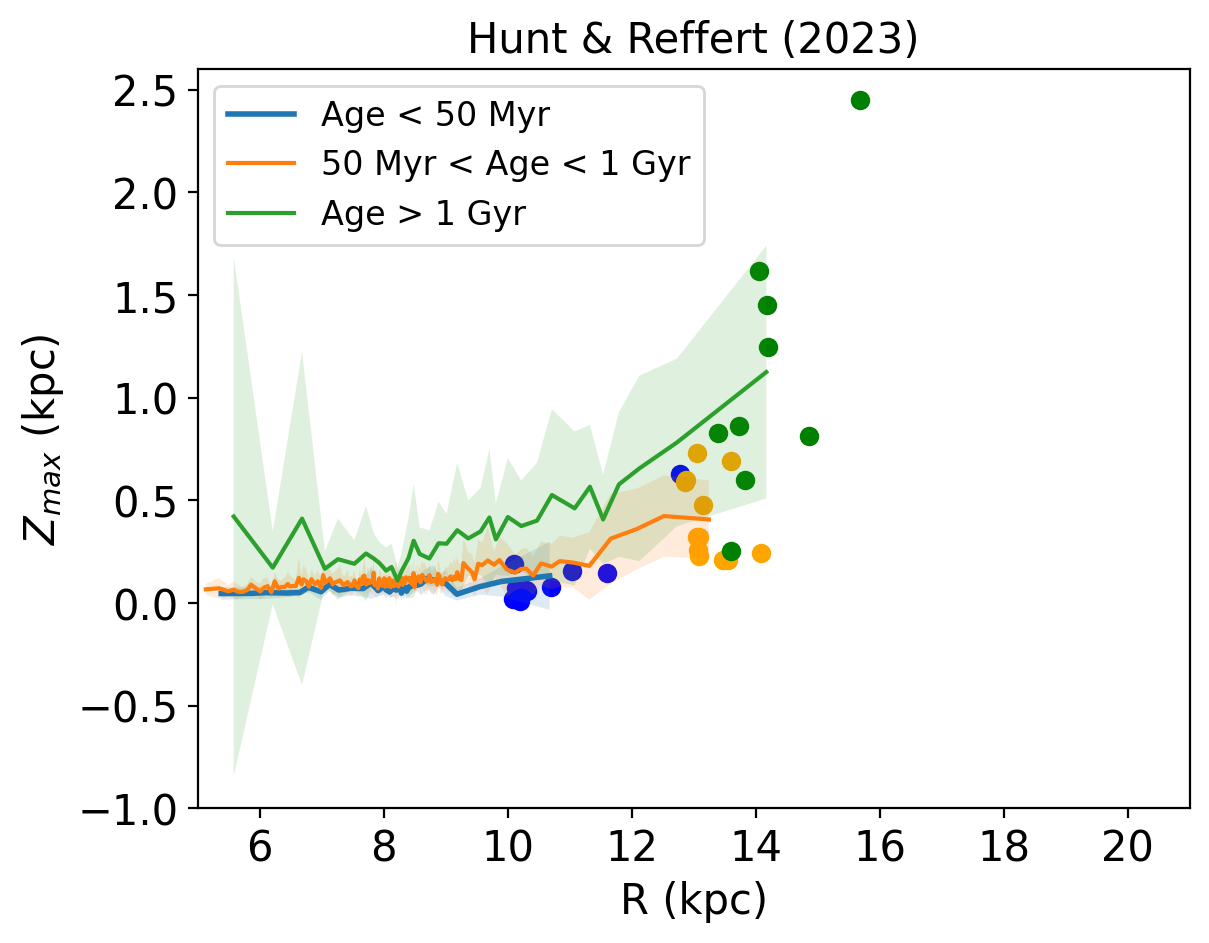}
    \qquad
    \caption{This figure presents a running mean for $Z_{max}$ values for every 20 consecutive OCs in current catalogue and 15 in \citet{2023A&A...673A.114H} for respective age bins, with their respective first sigma confidence intervals highlighted.
    }
    \label{all run mean}
\end{figure*}

We quantified these empirical observations by plotting the running mean plots for both catalogues, as shown in Fig. \ref{all run mean}. 
This figure shows the mean value of every 20 consecutive OCs in current catalogue and 15 consecutive OCs in HR23 and their 1-sigma confidence intervals for OCs in the respective age group (filled circles with respective colours represent the remaining OCs in each group). 
The profiles in the figure suggest that $Z_{max}$ is the lowest for young OCs, and the two older groups show larger values for the maximum vertical heights. 
We also note a distinct pattern in the $Z_{max}$ for the three age groups. 
The two younger age groups show a similar pattern up to a radius of 9 kpc, beyond which the intermediate age group show larger values for $Z_{max}$. 
The oldest age group also appears to have a differing pattern beyond 9 kpc, with the peak-like feature at a radius of 11 kpc. 
The pattern shown by the intermediate age and the old age group of the OCs suggests a flaring of the disk beyond a radius of 9 kpc.
Though the overall trend for the two catalogues is similar, there are differences in the details. 
HR23 has two peaks in the inner solar circle for the older population, which are not in current catalogue. 
The big jump after R $\sim$ 10 kpc is not visible in HR23, which is very prominent in current catalogue. 
There are many minor variations in the intermediate age population, possibly because the number of OCs is larger in HR23. 
The distribution is similar for the young population, mainly concentrated towards the disk mid-plane and has a small bump between 8 kpc $<$ R $<$ 9 kpc.

\begin{figure*}
    \centering
    \subfigure[]{\label{fit zmax age}
    \includegraphics[width=7.8cm]{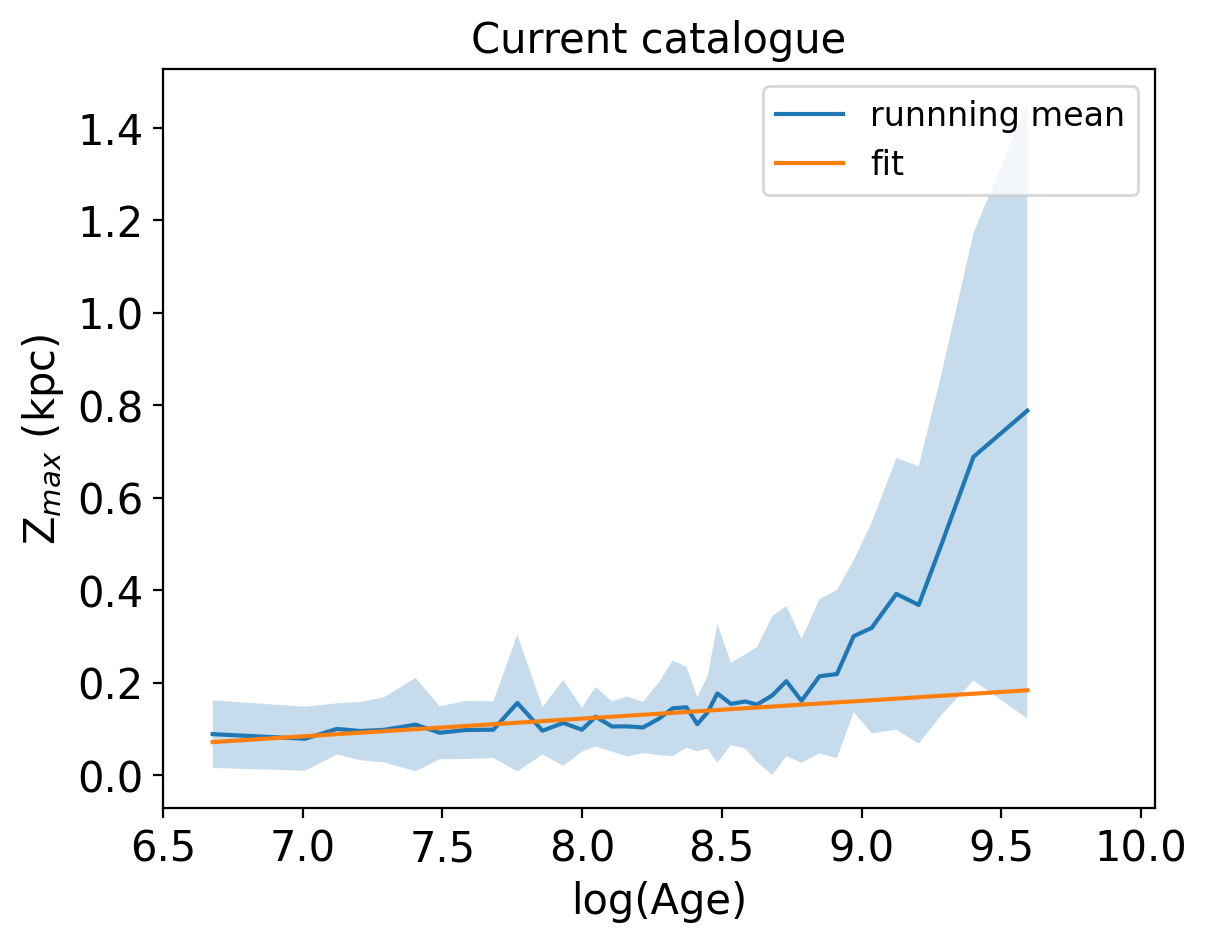}
    \includegraphics[width=7.8cm]{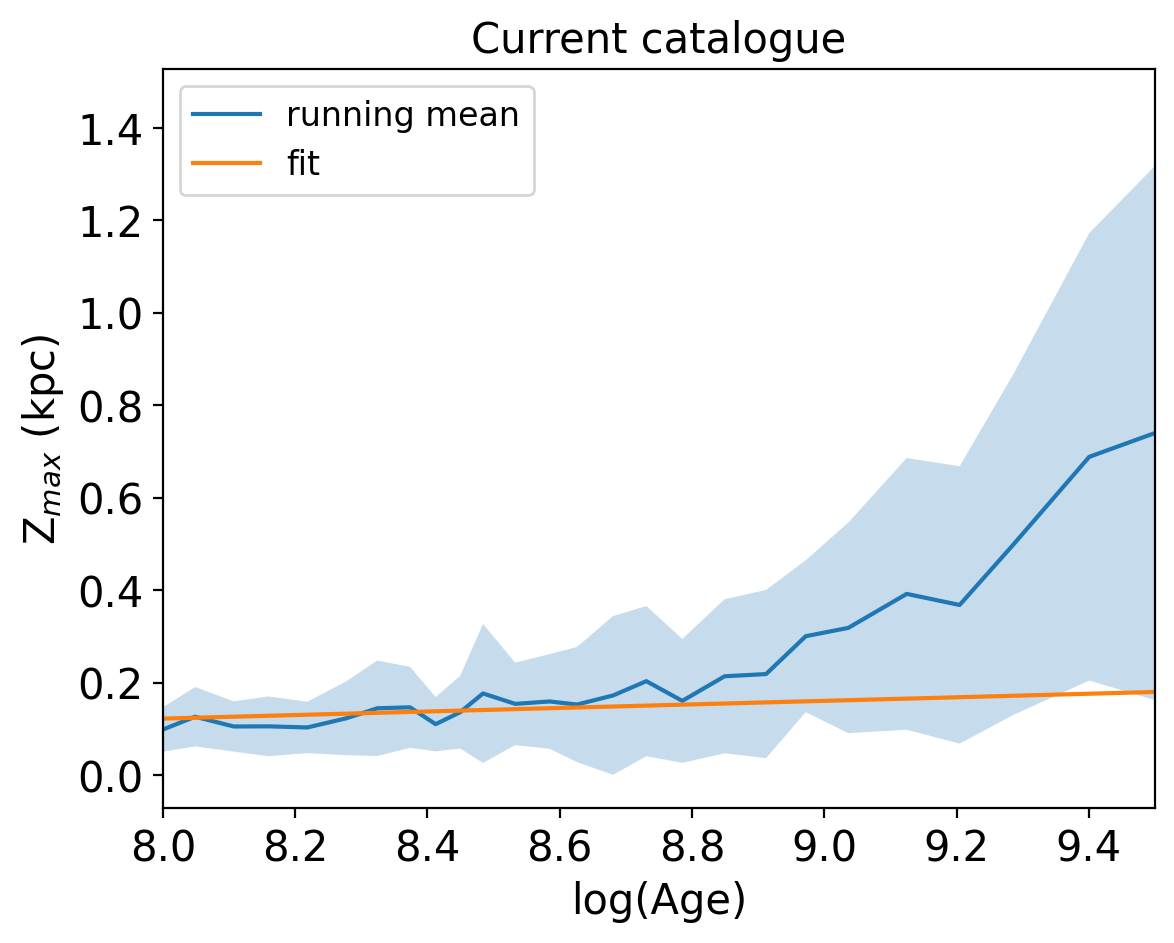}}
    \subfigure[]{\label{hunt fit zmax age}
    \includegraphics[width=7.8cm]{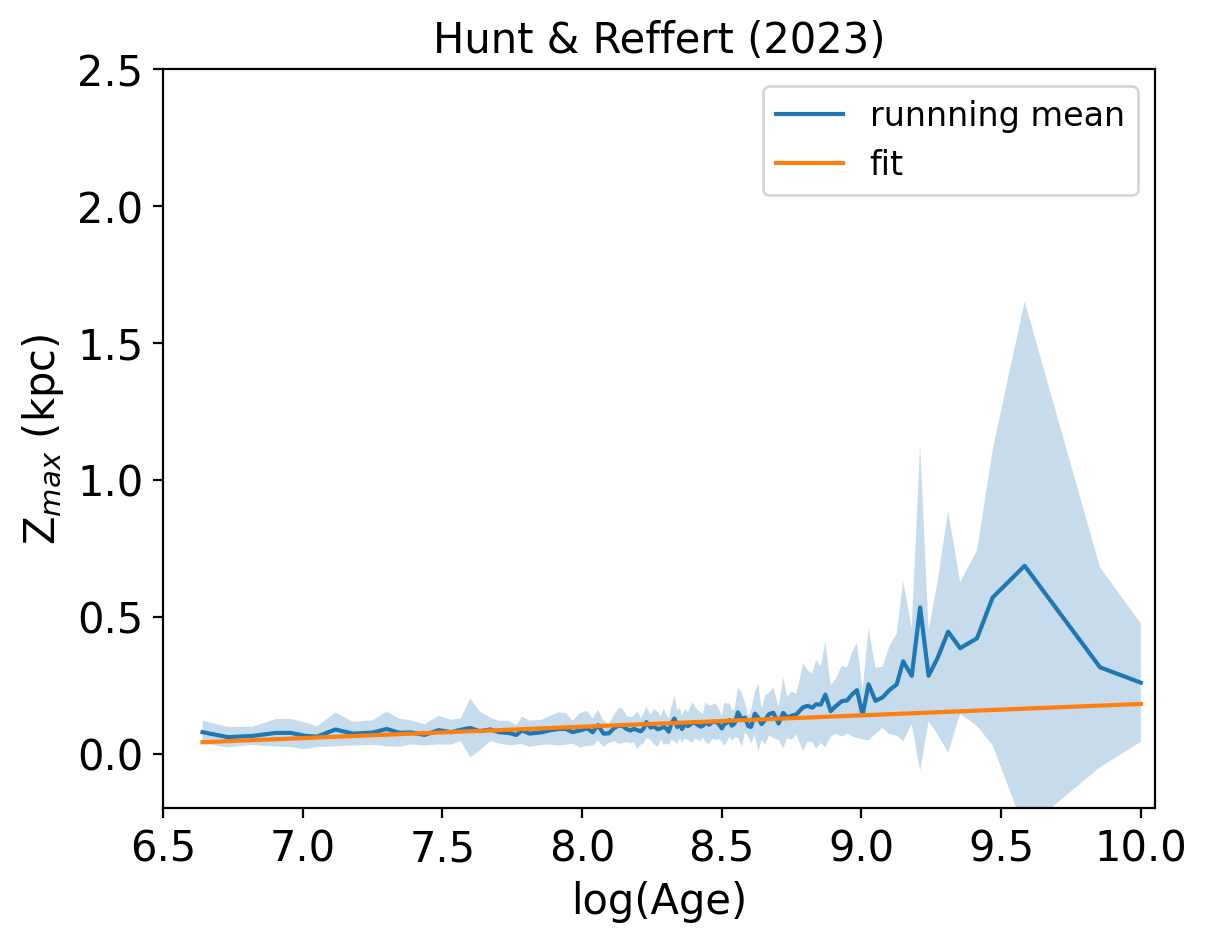}
    \includegraphics[width=7.8cm]{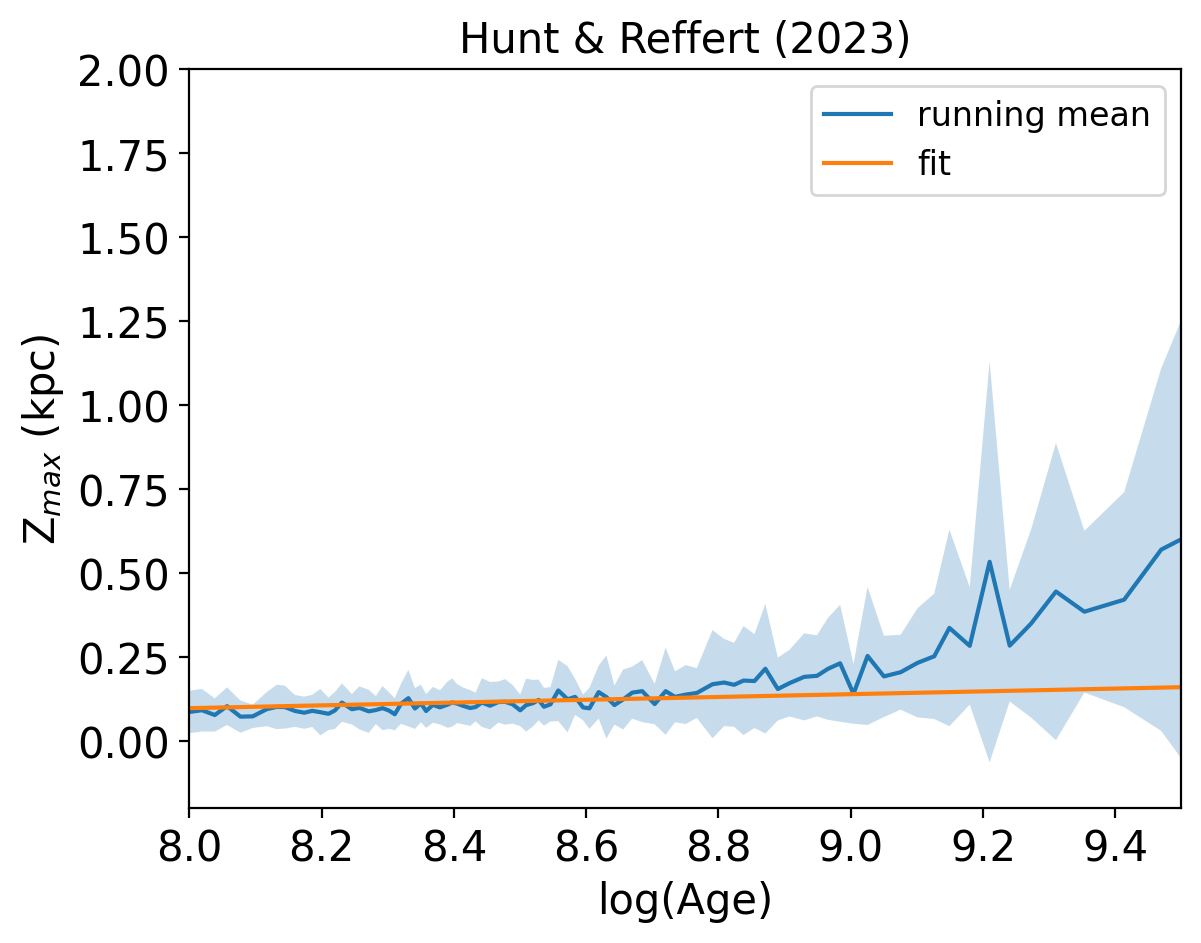}}
    \caption{The left panel of \ref{fit zmax age} presents the running mean (bin-width = 30) and first sigma confidence interval for OCs in current catalogue. A linear fit to the mean values of  $Z_{max}$ for OCs till log (Age)= 9 is also shown in orange to aid the eye and observe the cutoff cluster age, after which the vertical heights for OCs shoot up. And the right panel presents its zoomed-in perspective. \ref{hunt fit zmax age} shows similar figures for OCs in HR23.}
    \label{run mean}
\end{figure*}

\begin{figure*}
    \centering
    \subfigure[]{\label{fit zmax r}
    \includegraphics[width=7.8cm]{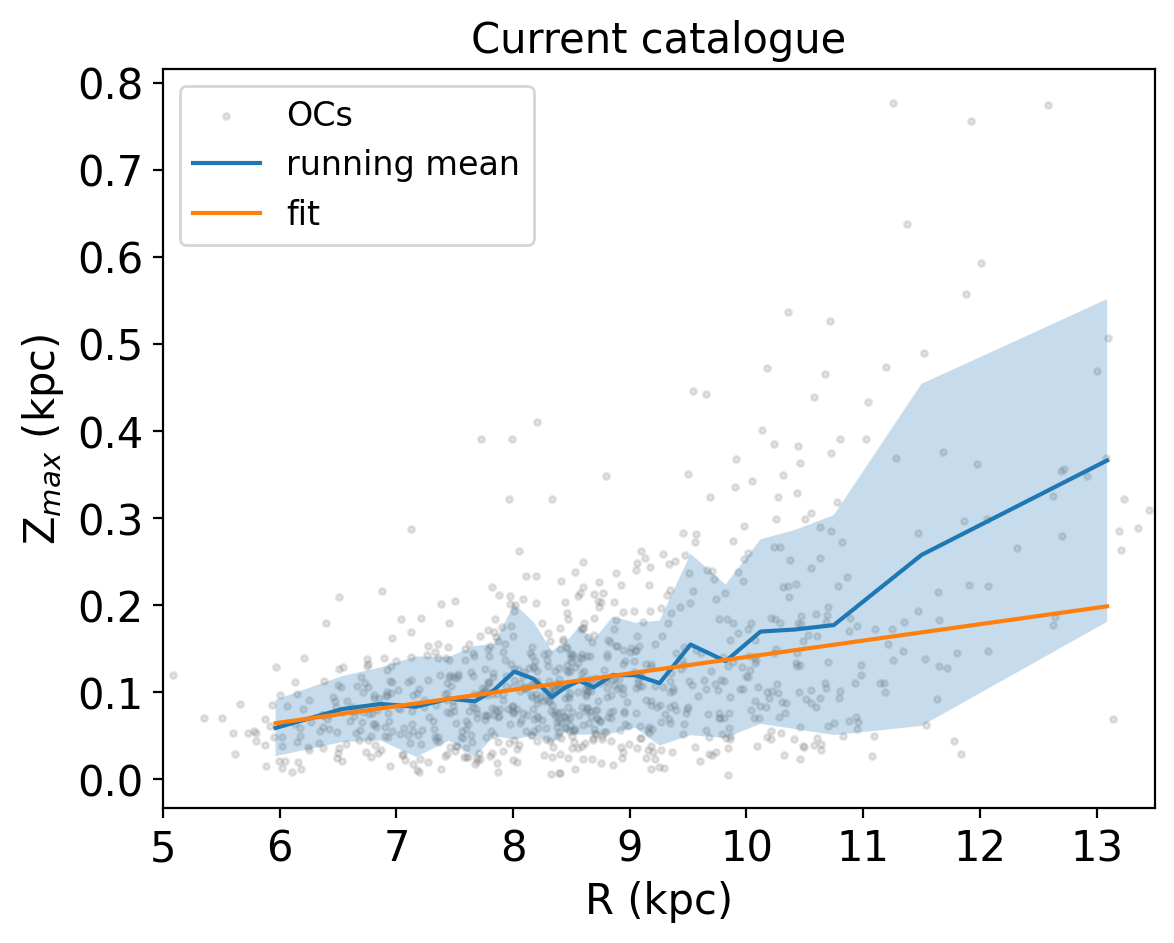}
    \includegraphics[width=8.2cm]{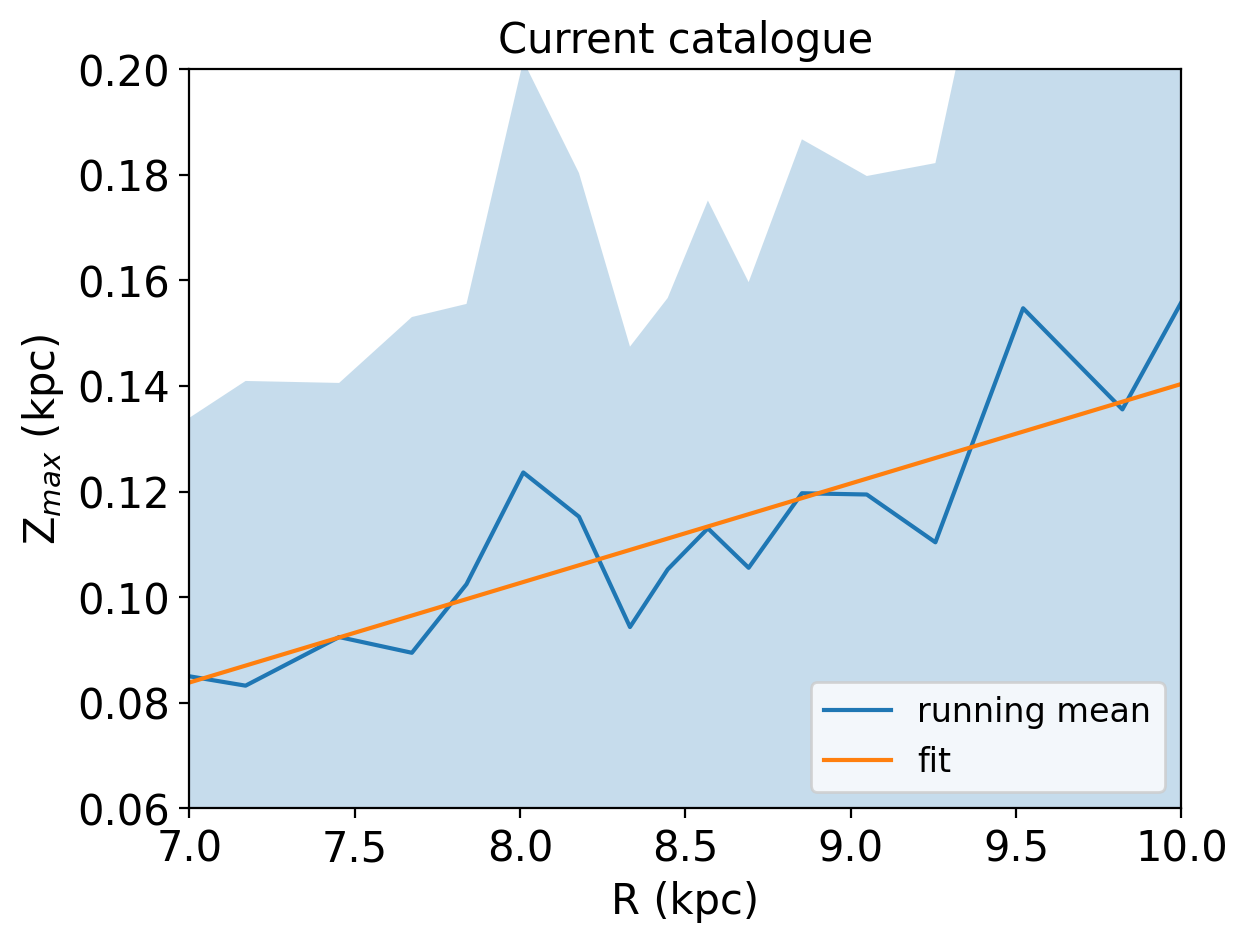}}
    \subfigure[]{\label{hunt fit zmax r}
    \includegraphics[width=7.8cm]{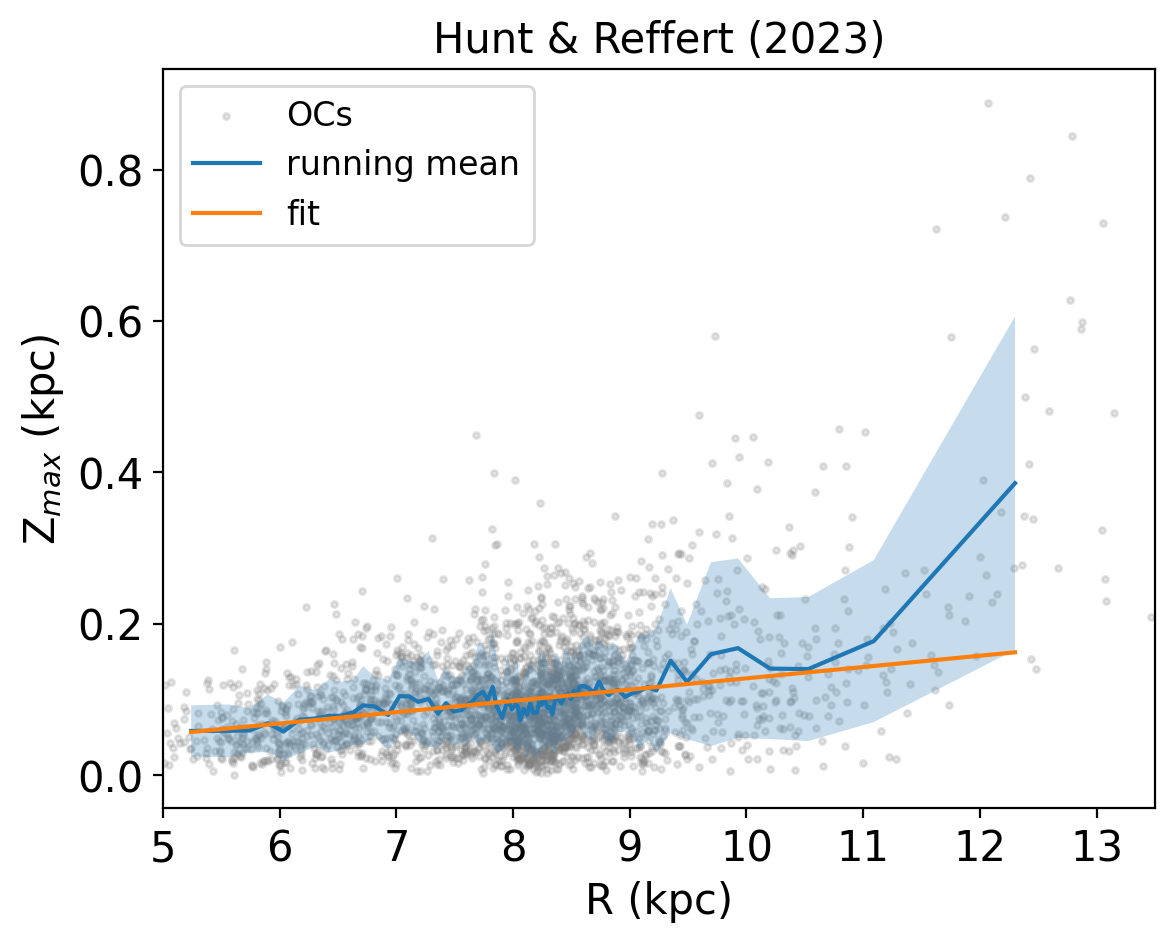}
    \includegraphics[width=8.2cm]{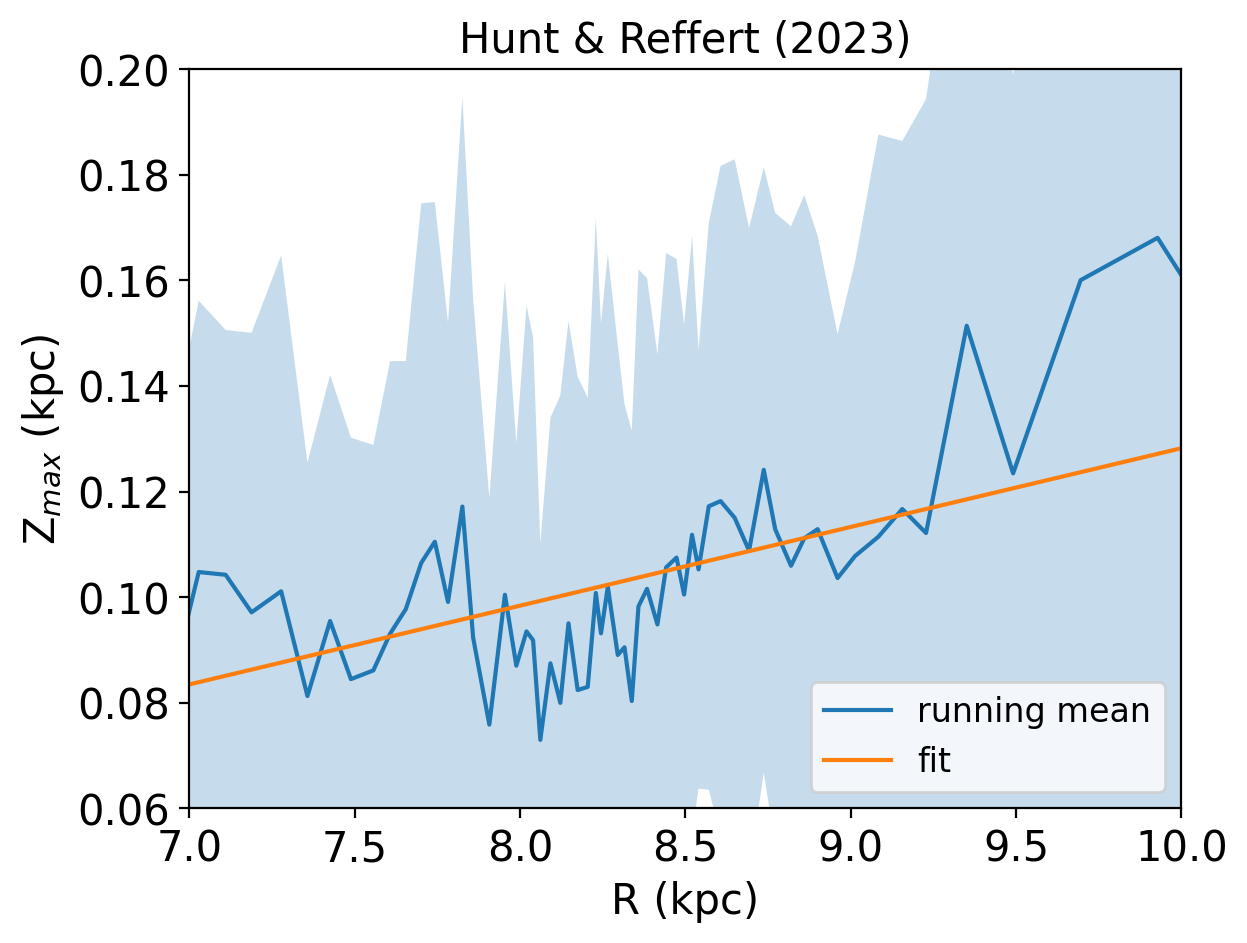}}
    \caption{Left panel of \ref{fit zmax r} presents the running mean and first sigma confidence intervals for OCs younger than 800 Myr in current catalogue. The orange line is the linear fit to the mean values of $Z_{max}$ of OCs within 10 kpc from the Galactic center. The right panel is a zoomed-in view of the left panel. OCs at a Galactocentric radius of 8 kpc and 9.5 kpc (and beyond) tend to reach greater vertical heights. \ref{hunt fit zmax r} present a similar plot for OCs in HR23.}
    \label{fits of zmax & R}
\end{figure*}

We note that the intermediate age group covers an extensive range of ages, and we attempt to identify the ages of the youngest OC that show flaring.
With this aim, we plotted the running mean for $Z_{max}$ values of the OCs with the cluster age in Fig. \ref{run mean}. 
We note that the profile shows relatively low values of $Z_{max}$ initially and a sharp increase for older OCs. 
We performed a linear fit on mean $Z_{max}$ values to trace the dependency of $Z_{max}$ on the age of the OCs. 
For OCs in the current catalogue, we estimated the following relation, $$ Z_{max} = (0.038\pm007) \times (log(age)) - (0.185\pm0.052) $$ where the applicability of this expression is upto the log (age) $\sim$ 9.0. For the first time, the above relation provides a quantitative value of $Z_{max}$ for various ages.
The fit also helps detect sudden jumps above and below the mean in vertical height, as shown in the figure. 
It appears that the deviations start around 1 Gyr. The right panel of Fig. \ref{fit zmax age} provides an enlarged view of a portion of the figure in the left panel, and we note that OCs older than $log (Age) = 8.9$ ($\simeq$ 800 Myr) tend to have larger values of $Z_{max}$, whereas the younger OCs have lesser values of $Z_{max}$.
In the case of OCs in HR23, a similar pattern is visible and with the following relation,  $$ Z_{max} = (0.041\pm004) \times (log(age)) - (0.234\pm0.029) $$ where the coefficients match within 1-sigma to the values obtained for current catalogue.

We estimated the median $Z_{max}$ values for various age bins of OCs in both catalogues.
For OCs younger than 10 Myr, median $Z_{max}$ is 57 pc (63 pc); in the age range 10 - 100 Myr, OCs reach a median vertical height of 90 pc (69  pc); OCs in age range 100 - 800 Myr reach a $Z_{max}$ of 113 pc (100 pc); whereas OCs older than 800 Myr have a median $Z_{max}$ of 339 pc (221 pc), based on the current catalogue (HR23 catalogue). 
We note that the value of $Z_{max}$ grows as a function of the cluster's age, but the values obtained from the current catalogue are larger than that from HR23. The difference is significant for the youngest and the oldest age groups. We note that the current catalogue has more OCs at large radii, contributing to a higher value of $Z_{max}$. 

To investigate the vertical pattern of higher $Z_{max}$ found with increasing $R$ in Fig. \ref{all run mean}, we plotted the running mean (bin width = 40) of $Z_{max}$ for OCs younger than 800 Myr as a function of $R$ in Fig. \ref{fits of zmax & R}.
For the first time, we estimate the dependency between $Z_{max}$ and $R$ as,
$$ Z_{max} = (0.019\pm002) \times R (kpc)- (0.048\pm0.020) $$ 
Left panels of Fig. \ref{fits of zmax & R} suggest that most OCs younger than 800 Myr reach maximum vertical heights not greater than 200 pc within a $R$ of 10 kpc. 
In contrast, some OCs achieve vertical heights of greater than 200 pc beyond a radius of 11 kpc, comparable to the increased value of $Z_{max}$ for the OCs older than 1 Gyr (see fig.8). 

The right panels present an enlarged part between the $R$ of 7 - 10 kpc. 
In this running average profile, two prominent vertical features are visible for OCs in the current catalogue.
We note that at two $R$ values (8 kpcs and 9.5 kpc), $Z_{max}$ shoots up by nearly 20-30 pc. This increased value of $Z_{max}$ has a width not more than 500 pc at the values as mentioned earlier of the radii.
One could also naively assume that these vertical features might correlate to the spiral structure of the Galaxy and that the tidal forces of the dense spiral arms might perturb the OCs' vertical heights.

The dependence of $Z_{max}$ as a function of R is observed in the case of HR23, with a very similar slope and a slightly different value for the y-intercept. 
The relation is found to be:
$$ Z_{max} = (0.0149\pm001) \times R (kpc) - (0.021\pm0.011) $$

The relation from HR23 suggests a similar variation in $Z_{max}$ with radial distance with respect to the current catalogue.
The prominent peak at R $\sim$ 8 kpc observed in the OC distribution of the current catalogue is shifted inwards ( between 7.5 kpc and 8 kpc) in the distribution of OCs from the HR23 catalogue. We note that the number of OCs in a bin is the same in both catalogues. Therefore,
there are more number of bins in HR23 as the number of OCs is larger in HR23, resulting in several peaks in HR23 OC distribution in comparison to a few peaks seen in the current catalogue OC distribution.
The peak at R $\sim$ 9.5 is clearly visible in both catalogues, whereas we note another peak just beyond in HR23. The calculated confidence level of these three peaks is 99\%, whereas the other visible peaks have a confidence level of less than 95\%. Overall, both catalogues provide similar inferences on the radial distribution and related features of OCs.

\begin{figure*}
    \centering
    \subfigure[]{
    \label{sp arms vs. radius}
    \includegraphics[width=0.49\textwidth]{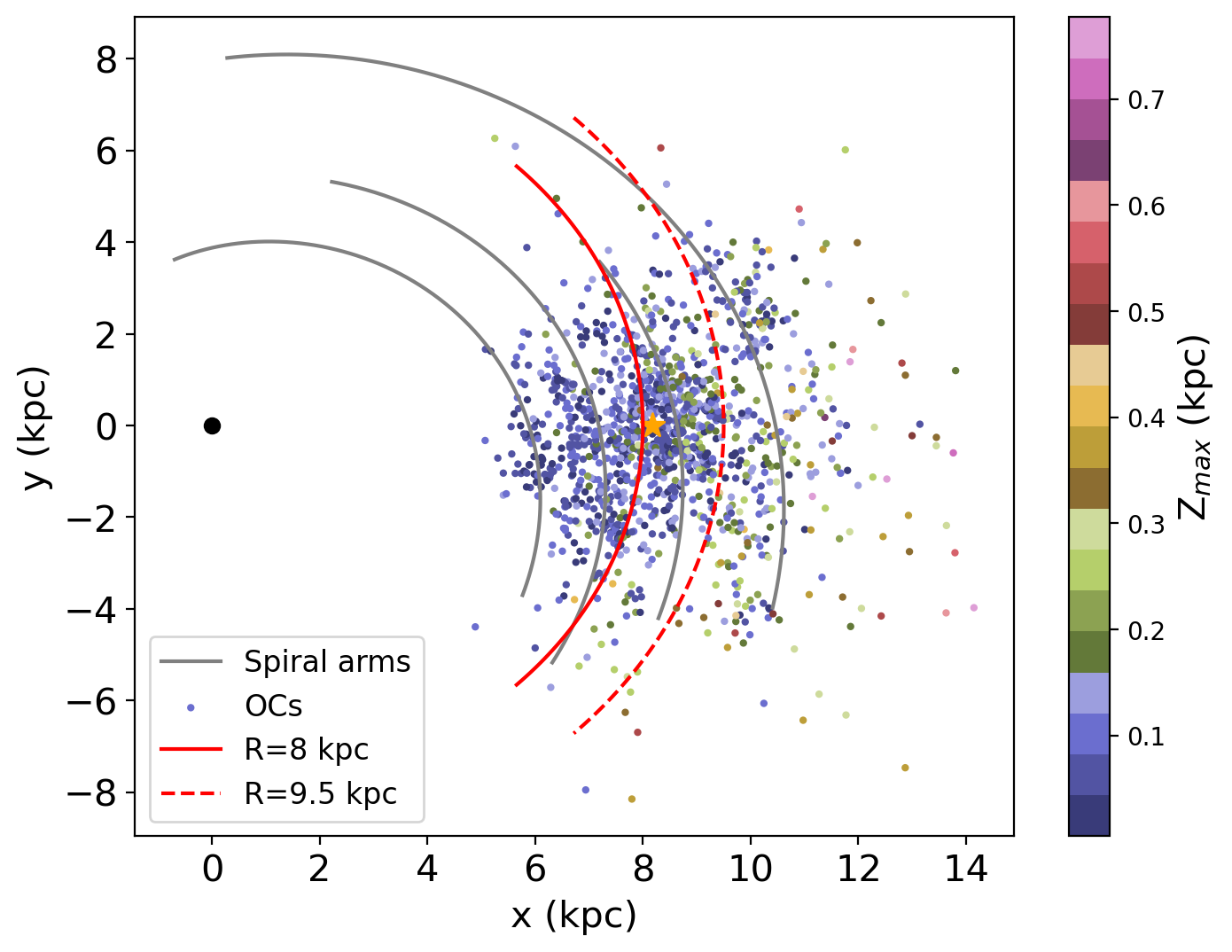}}
    \hfill
    \subfigure[]{
    \label{hist rgc}
    \includegraphics[width=0.49\textwidth]{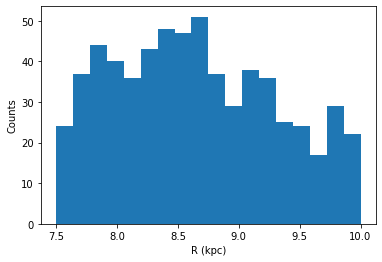}}

    \subfigure[]{
    \label{kde l Zmax}
    \includegraphics[width=.49\textwidth]{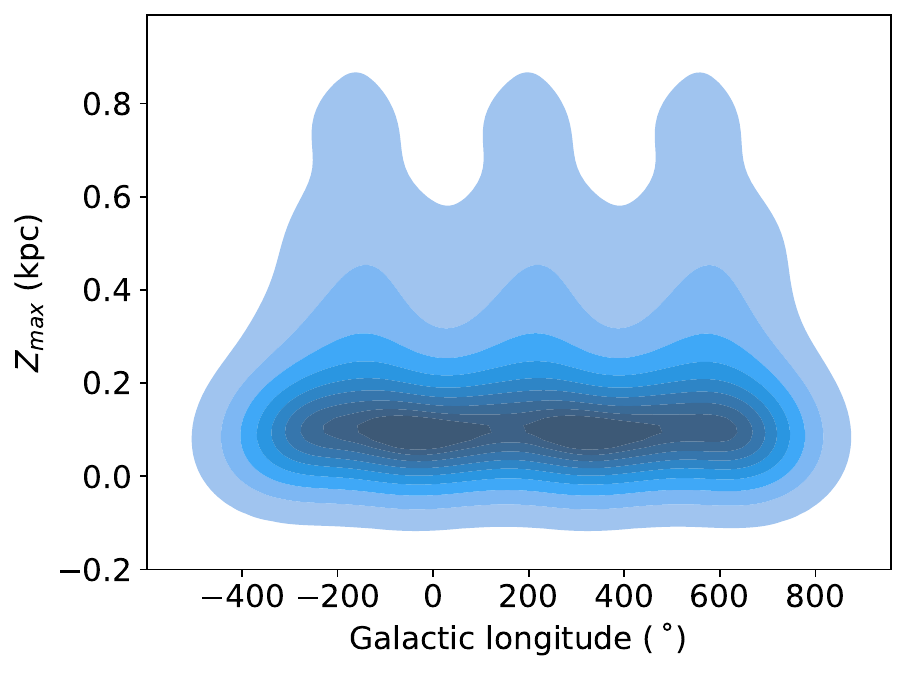}}
    
    \qquad
    \caption{Fig. \ref{sp arms vs. radius}: OCs younger than 800 Myr taken from the current catalogue, in cartesian Galactocentric coordinates (colour-coded to their respective maximum vertical heights $Z_{max}$). The position of the Sun is marked by an orange star at (8.178,0), and the black dot represents the Galactic center. Spiral arms are fitted using the parameters derived in \citet{2021A&A...652A.162C}. Spiral arms from the Galactic center towards the Sun are in order: Scutum-Centaurus arm, Sagittarius-Carina arm, Local arm, and the farthest is the Perseus arm. Here, the curves for radius 8 kpc and 9.5 kpc centred at the Galactic center are marked by red lines. Fig. \ref{hist rgc}: A histogram across $R$ for OCs younger than 800 Myr is shown here. A higher number density is seen in the regions near the local arm. In contrast, the OC population at radial distances of 8 kpc and 9.5 kpc is comparatively scarce. \ref{kde l Zmax} represents the KDE in longitude and $Z_{max}$ plane for all the OCs in our analysis.}
    \label{Fig:sp}
\end{figure*}

\begin{figure*}
    \centering
    \subfigure[]{
    \label{sp_hunt}
    \includegraphics[width=0.49\textwidth]{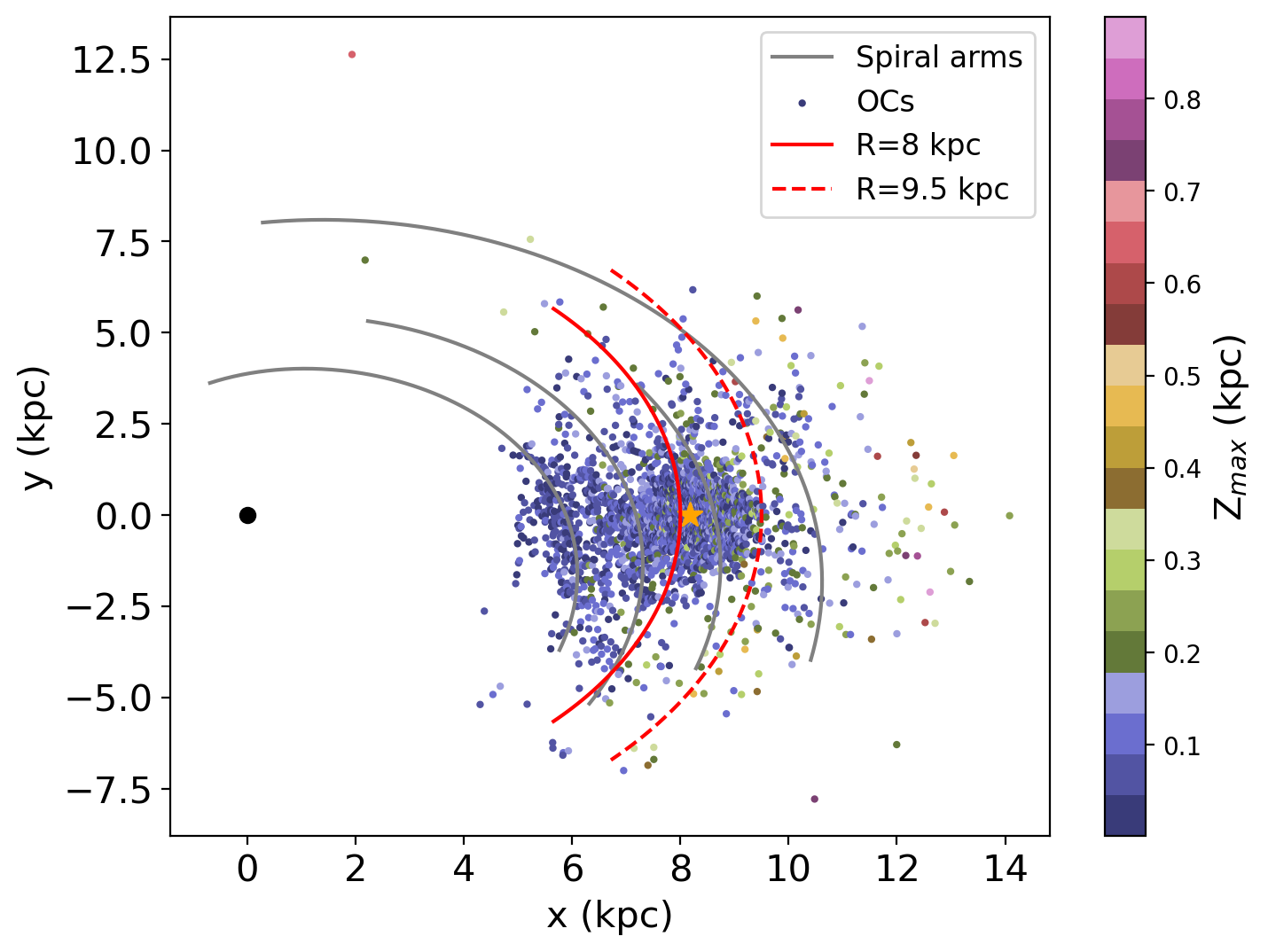}}
    \hfill
    \subfigure[]{
    \label{hist_Rhunt}
    \includegraphics[width=0.49\textwidth]{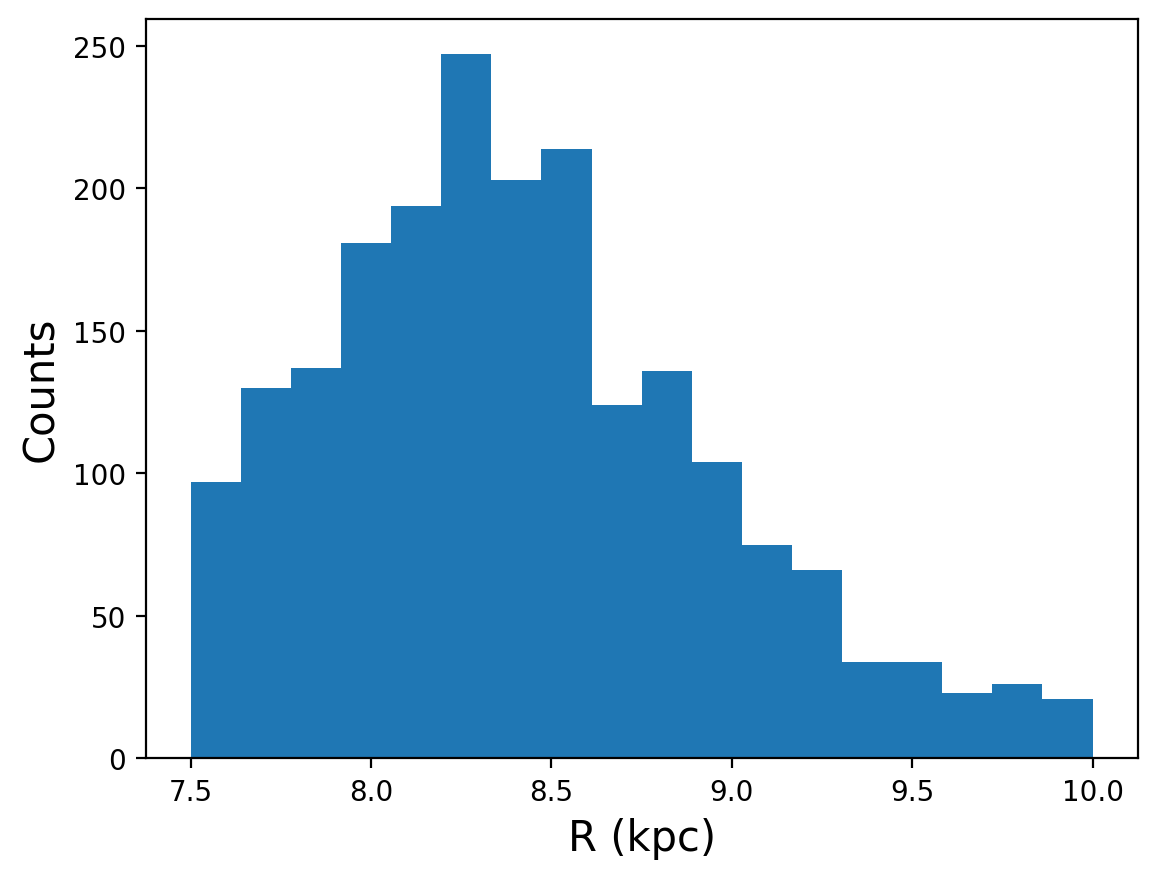}}

    \subfigure[]{
    \label{kde_hunt}
    \includegraphics[width=.49\textwidth]{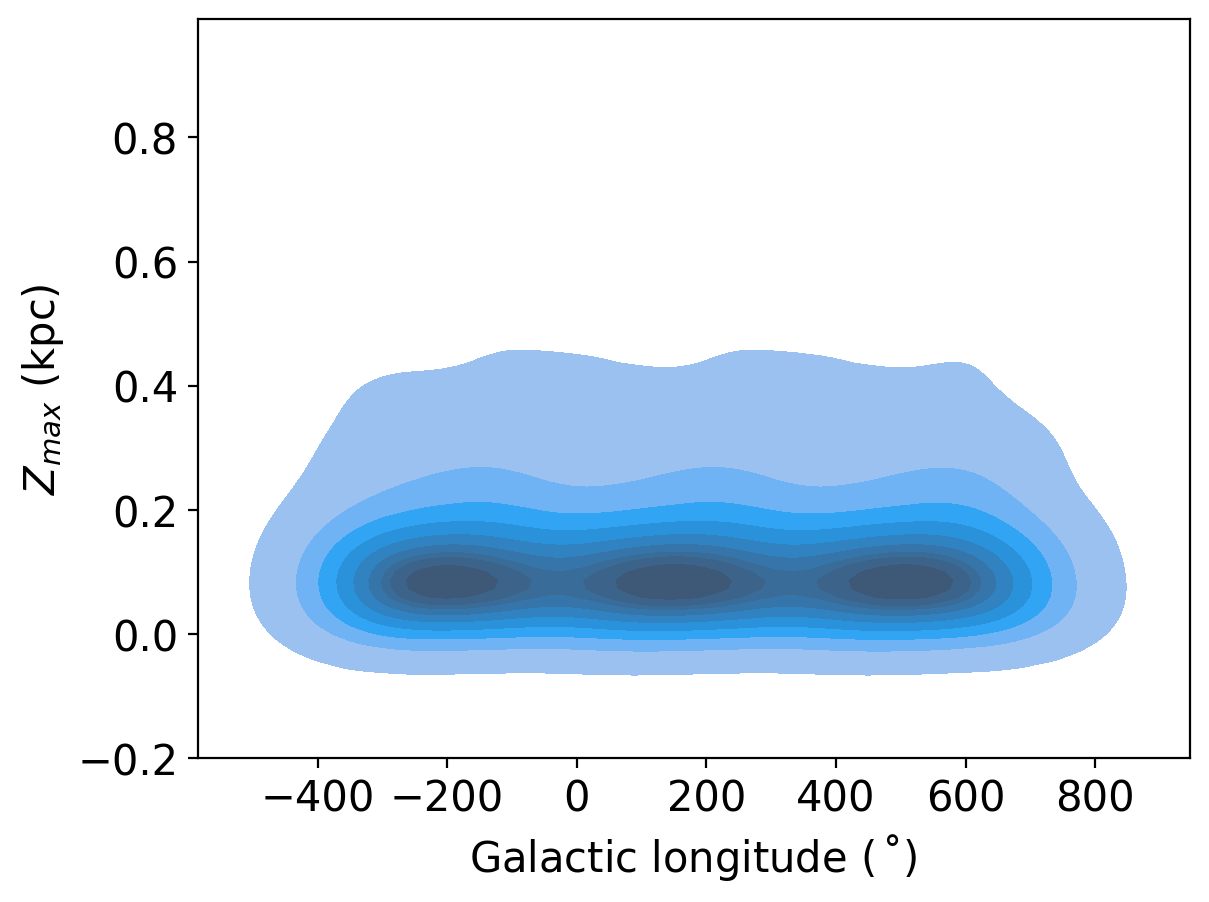}}
    
    \qquad
    \caption{A similar figure as \ref{Fig:sp} for the OCs in HR23.}
    \label{sp}
\end{figure*}

To explore the relationship between the increased values of $Z_{max}$ and spiral arms, we have shown the locations of the OCs from the current catalogue in the X-Y plane (Fig. \ref{sp arms vs. radius}). We also plotted the locations of spiral arms and the (partial) circles corresponding to the $R$, 8 and 9.5 kpc. 
The OCs younger than 800 Myr (colour-coded with their $Z_{max}$) are also shown here, with the position of the Sun marked within the Galactic plane by an orange star at 8.178 kpc.  
Interestingly, the arcs corresponding to the two radii of interest majorly occupy regions between the spiral arms. 
Therefore, this figure suggests that OCs tend to have larger $Z_{max}$ values in the inter-arm region.
In Fig. \ref{hist rgc}, the number of OCs as a function of $R$ is shown. We note that the largest number of OCs is around 8.5 kpc, aligning with the Local spiral arm. We do not detect any significant peaks at 8 kpc and 9.5 kpc, suggesting that regions with a localised increase in $Z_{max}$ are away from the spiral arms. 

In order to assess the correlation between $Z_{max}$ and the Galactic longitude ($l$), we have shown the KDE between $l$ and $Z_{max}$ as shown in Fig. \ref{kde l Zmax}. To reproduce the periodicity, we replicated the data points for l $<$ 0 and l $>$ 360.
This figure shows that the OCs between $l \sim 90$ and $l \sim 320$ are orbiting with an increased $Z_{max}$, with peak between $l \sim 200-250 $. 
This indicates the presence of $Z_{max}$ only in the outer Galactic disk.

We performed a similar analysis for HR23, and the distributions are shown in Fig. \ref{sp}.
A clear increase in the OC population in the solar neighborhood is visible in Fig. \ref{sp_hunt} that is causing a prominent peak between 8 kpc $<$ R $>$ 8.5 in Fig, \ref{hist_Rhunt}. 
In this histogram also, there are no peaks around 8 kpc and 9 kpc, which is a similar trend to the histogram in \ref{hist rgc}.
We note two peaks at l $\sim$ 120 $\deg$ and l $\sim$ 250 $\deg$ in Fig. \ref{kde_hunt}. The peaks visible in Fig. \ref{kde l Zmax} more or less match those in  Fig. \ref{kde_hunt}. 

\begin{figure*}
    \centering
    \includegraphics[width=8cm]{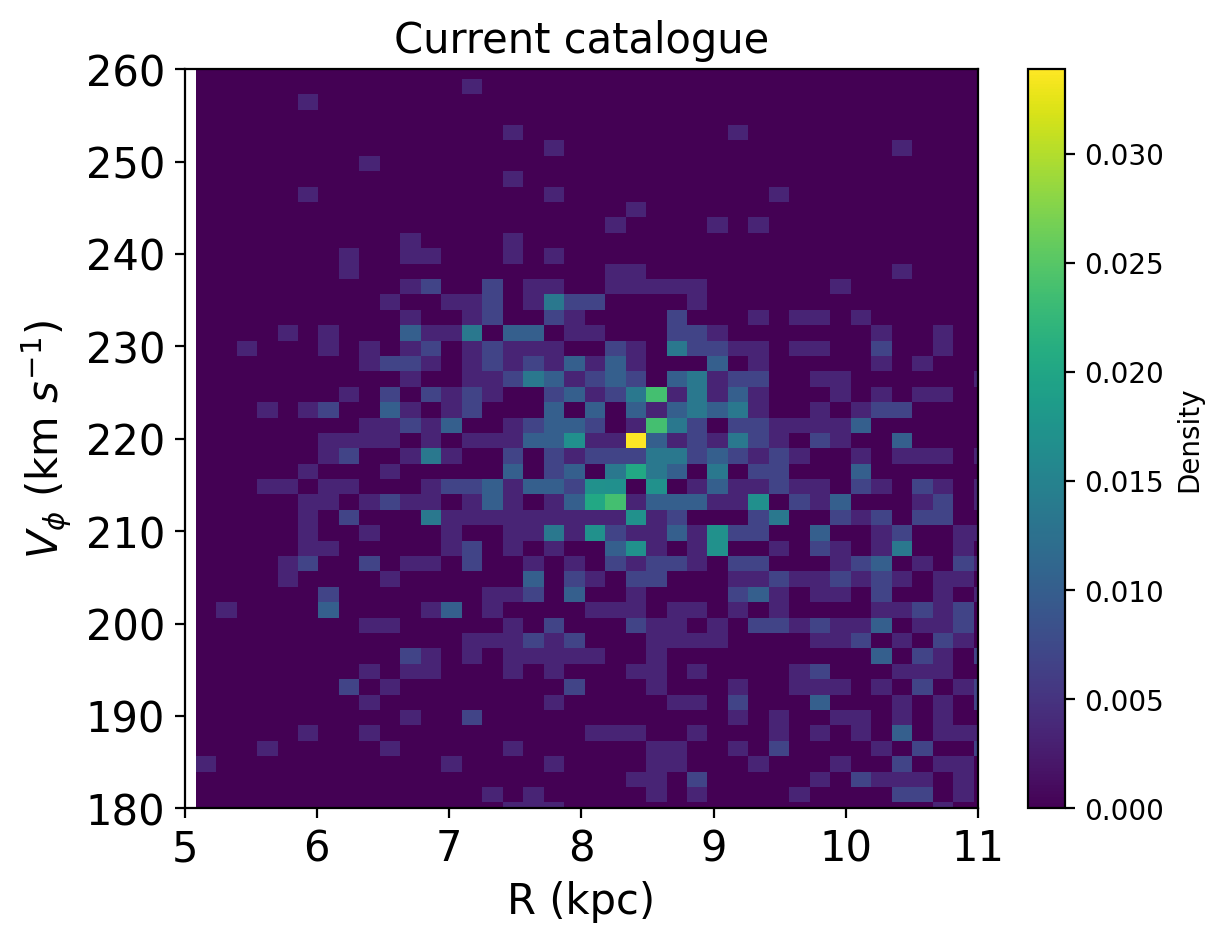}
    \includegraphics[width=8cm]{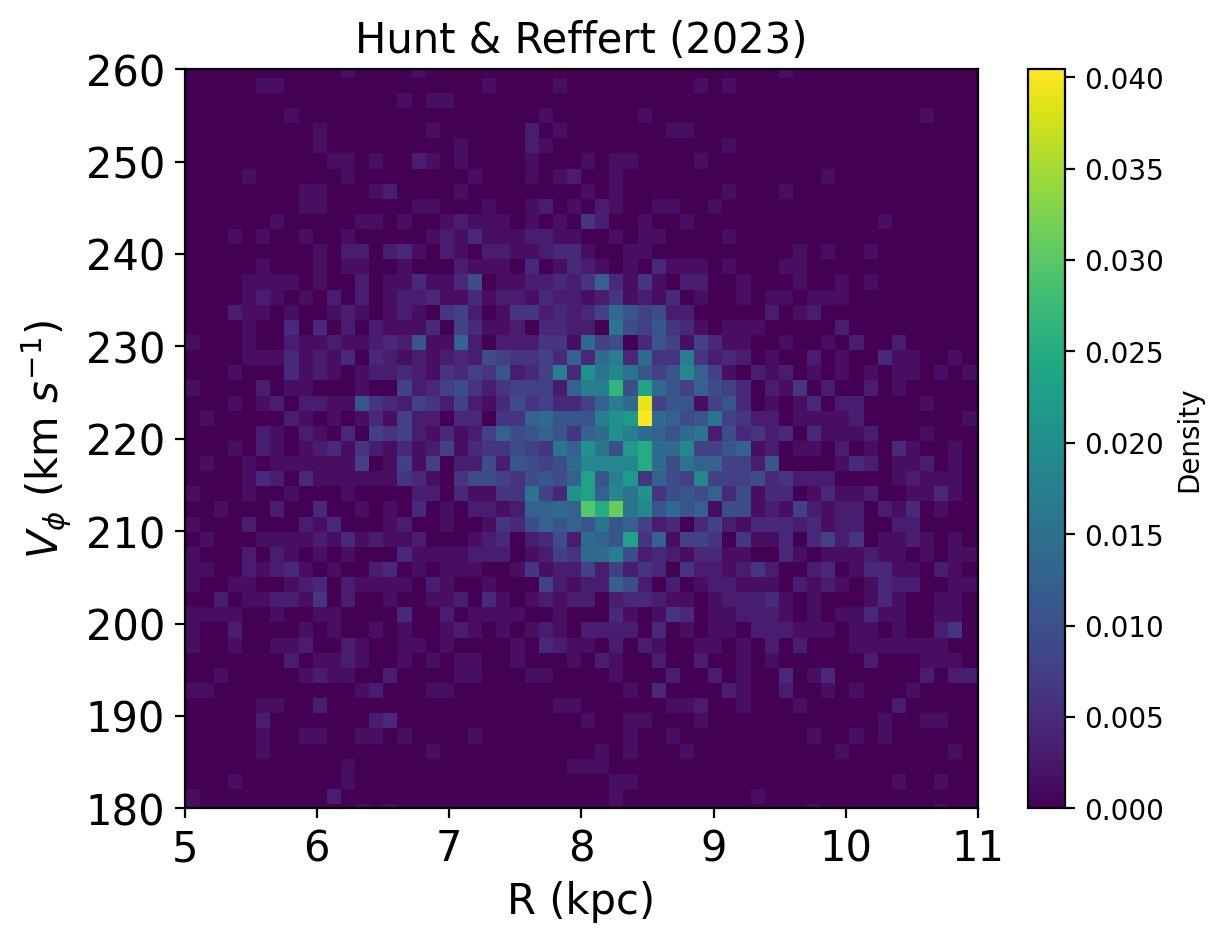}
    \qquad
    \caption{The distribution of the azimuthal velocities of the OCs from both catalogues as a function of their $R$. The plot for \citet{2023A&A...673A.114H} confirms the presence of the diagonal ridges in the solar neighbourhood as mentioned by \citet{2018Natur.561..360A}.
    }
    \label{ridge}
\end{figure*}

We also looked for the presence of diagonal ridges in the solar neighborhoods as found by \citet{2018Natur.561..360A}.
For this, we plotted the azimuthal velocity (V$_{\phi}$) of the OCs as a function of their $R$, as shown in Fig. \ref{ridge} for both catalogues.
This figure suggests that the ridges are not visible in current catalogue, but they are clearly visible in HR23.
\citet{Tarricq_2021} also showed the involvement of OCs in ridge formation using a sample of 1382 OCs from Gaia DR2.
\citet{2019MNRAS.489.4962K}, using the N-body simulation, showed that the stars located close to the Galactic midplane and having solar metallicity are involved in the prominent ridges.
Fig. \ref{ridge} also shows that the OCs involved in the ridge formation are orbiting at very low distances from the Galactic midplane.
So, the thin-disk stars contribute to the ridge formation and support the result by \citet{2019MNRAS.489.4962K}.
The ridges shown in Fig. \ref{ridge} also follow the pattern shown by the field stars, which is the ridges start fading out for the larger $R$.

\section{Discussion}\label{sec:discussion}

The first contribution of this study is the catalogue of 1145 OCs with updated parameters from Gaia DR3. 
This will be an instrumental catalogue for various purposes, such as studying the OC population and exploring the properties of the Galactic disk. 
In this study, using the kinematic information derived from the current catalogue for 1145 OCs along with 3677 OCs from HR23, we estimated the orbits of 4006 OCs. 
This is used to study the orbital motion of OCs and their vertical distribution in the Galactic disk.
Our sample of OCs is distributed within a Galactocentric radial distance of 5 - 16 kpc, and the properties of the disk explored in this study pertain to this range.
We note that the OC sample is based on data availability and, therefore, can be incomplete. 
The results presented in this study can reflect a bias towards nearby regions due to the unavailability of data in the other parts of the Galactic disk.

In general, the distribution of OCs perpendicular to the Galactic plane measures the scale height or thickness of the Galactic disk. 
The value of scale height is usually estimated from the distribution of the observed $Z$ values of OCs \citep{2006A&A...446..121B, 2023AJ....166..170J}. 
This $Z$ value will always be less than the actual height of the orbits taken by OCs in the Galaxy. 
In this study, we trace the actual height the OCs can achieve.
The maximum height achieved by any OC in its orbit depends on its position in the disk. 
The OCs located in the inner part are orbiting vertically closer to the Galactic mid-plane. While, the OCs located in the outer regions of the Galactic Disk are found to gain greater vertical heights. 
A similar trend was found by \citet{2006A&A...446..121B} as the scale height doubled for OCs outside compared to those inside the solar circle. 
\citet{2014MNRAS.444..290B} and \citet{2016A&A...593A.116J} found the scale height to increase with Galactocentric radius, and \citet{2023AJ....166..170J} further confirmed the trend.
\citet{2024arXiv240614661M} showed disc thickening using OCs. 
They concluded that the disc thickening is a consequence of disruption of OCs close to the Galactic mid-plane due to interaction with the disc.
In this study, we quantified this trend by making a linear fit to $Z_{max}$ vs Galactocentric radius estimation. The fit also exposed an overall flaring up of the disk beyond 10 kpc across all age groups, primarily for old OCs. The disk flaring with the $R$ was also observed by \citet{2014ApJ...794...90K, 2014A&A...567A.106L, 2016ApJ...823...30B}.

A trend between $Z_{max}$ and cluster age was also found because a greater population of older OCs tends to gain higher vertical heights in their trajectories around the Galactic center.
The maximum vertical distance travelled by the younger OCs is limited to a few hundred parsecs, whereas the older OCs tend to achieve 3 to 4 times the $Z_{max}$ of younger OCs. 
We derive a relationship between $Z_{max}$ and log(age) of the cluster and bring out the fact that the relation deviates for OCs older than log(age)=8.9. 
There are estimations of scale height for different age groups and indications that the scale height increases for older OCs in the literature such as 74 $\pm$ 5 pc for OCs of 100 Myr by \citet{2020A&A...640A...1C},  70.5 $\pm$ 2.3 pc for OCs younger than 20 Myr and 87.4 $\pm$ 3.6 pc for 20-100 Myr OCs by \citet{2021A&A...652A.102H}, and   91.7 $\pm$ 1.9 pc for OCs younger than 700 Myr by \citet{2023AJ....166..170J}. 
%This study finds that the median value of $Z_{max}$ for OCs younger than 10 Myr is 60 pc, 10-100 Myr is 90 PC, and 100-800 Myr is 110 PC. 
In the case of OCs older than 800 Myr, the median value of $Z_{max}$ shoots to 221 pc. 
 
The value of $Z_{max}$ is a function of both age and Galactocentric radius and shows a positive trend. 
Over and above this dependency, we detect specific jumps in $Z_{max}$ for OCs older than log(age) = 8.9 and OCs located beyond 10 kpc, along with two noticeable bumps at 8 and 9.5 kpc. 
Recently, \citet{2024MNRAS.527.4863U} found evidence for warp and flare of the old Galactic disk as traced by red clump stars. 
They found the scale height of the disk to range from 380 pc in the Solar neighbourhood to 2.5 kpc at a Galactocentric radius of 15 kpc. 
In this study, the highest value of $Z_{max}$ is found to be 2.75 kpc.
 
We noticed that OCs, specifically in the inter-arm regions of the disk, have higher $Z_{max}$ values than OCs near 
or within the spiral arms. 
We also checked for the possibility of an observation bias and found that the number density of OCs at a distance of 8.0 and 9.5 
kpc from the galactic center is lower than other regions of the $R$ (the highest being just before the local arm at a radius of 8.3 kpc from the Galactic center).
 This observed jump in Z-direction at a specific Galactocentric distance may correlate with the newly discovered phase space spiral \citep{2023A&A...673A.115A}. 
We have not explored this aspect in this study, but it could be done in the future.

The patterns identified in this analysis might relate to the already known warps in the Galactic Disk. 
We then investigated this possibility and found a higher $Z_{max}$ in the second and third Galactic quadrants. 
Studies by \citet{2002A&A...394..883L, 2006A&A...451..515M, refId0} and \citet{2023ApJ...954L...9H} found Galactic warp in the same region of the Galaxy. 
This coincidence may suggest that due to the presence of warp in the galactic disk, the OCs are able to orbit at a higher distance from it. 

The overall nature of the results obtained using the sample of OCs from current catalogue and HR23 are similar, with a small difference in the minute details. 
Both samples show disc flaring in the outer regions, which is mainly caused by the older population of OCs.
The sudden jump in the $Z_{max}$ at R $\sim$ 9.5 kpc is visible in both samples, while the jump at R $\sim$ 8 kpc observed for OCs in the current catalogue is not so prominent for OCs in HR23.
Compared to the current catalogue, most of the new OCs included in the HR23 are located in the solar neighbourhood, hence giving rise to OCs between 8 $< R <$ 8.5 kpc.
The number of OCs at R $\sim$ 9.5 are similar in both catalogues, showing the same distribution.
The distribution of maximum height attained by the OCs as a function of the Galactic longitude is also more or less similar.
The diagonal ridges in the solar neighbourhood are visible using the OCs from HR23, suggesting that the younger OCs contribute to ridge formation.
This is consistent with the results obtained by \citet{2019MNRAS.489.4962K} using N-body simulations.

\begin{figure*}
    \centering
    \includegraphics[width=0.49\linewidth]{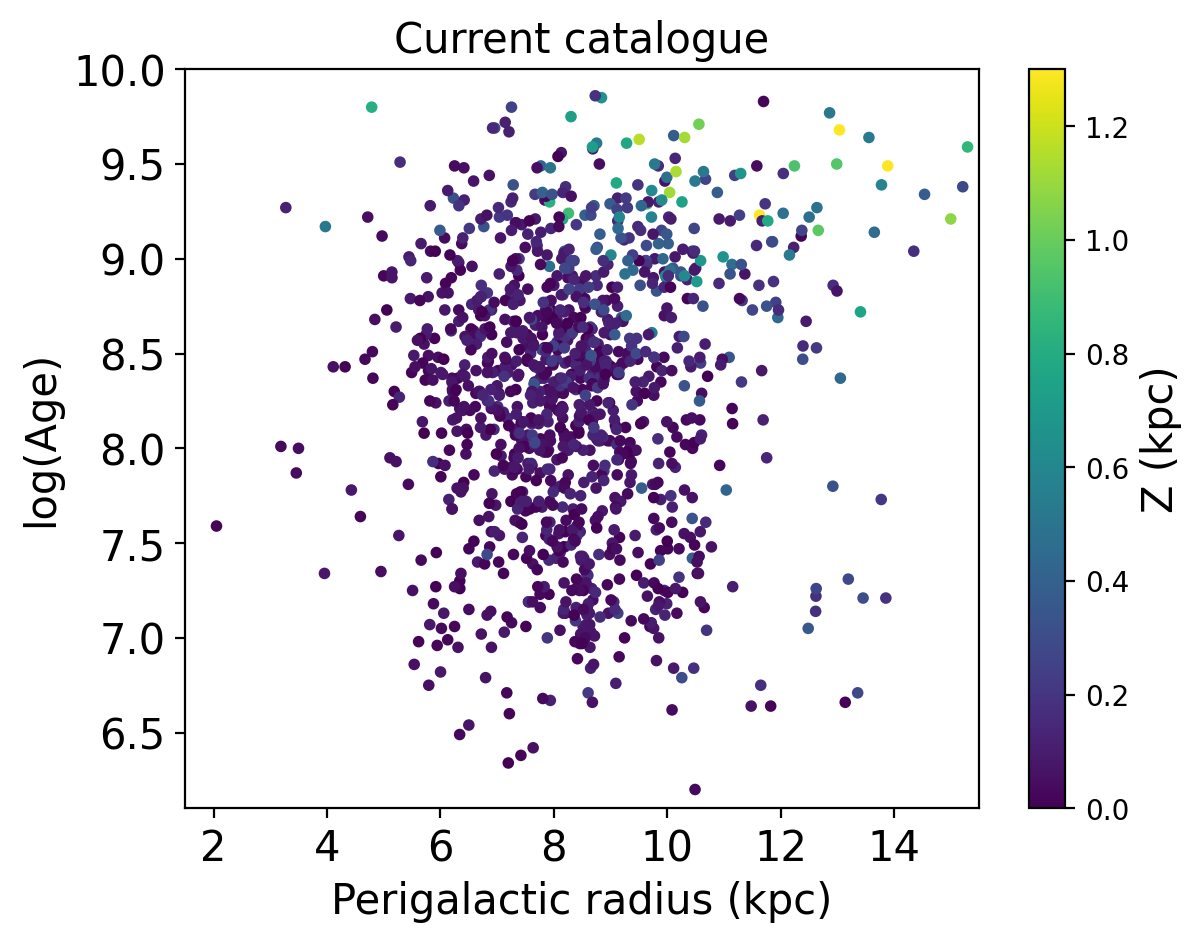}
    \includegraphics[width=0.49\linewidth]{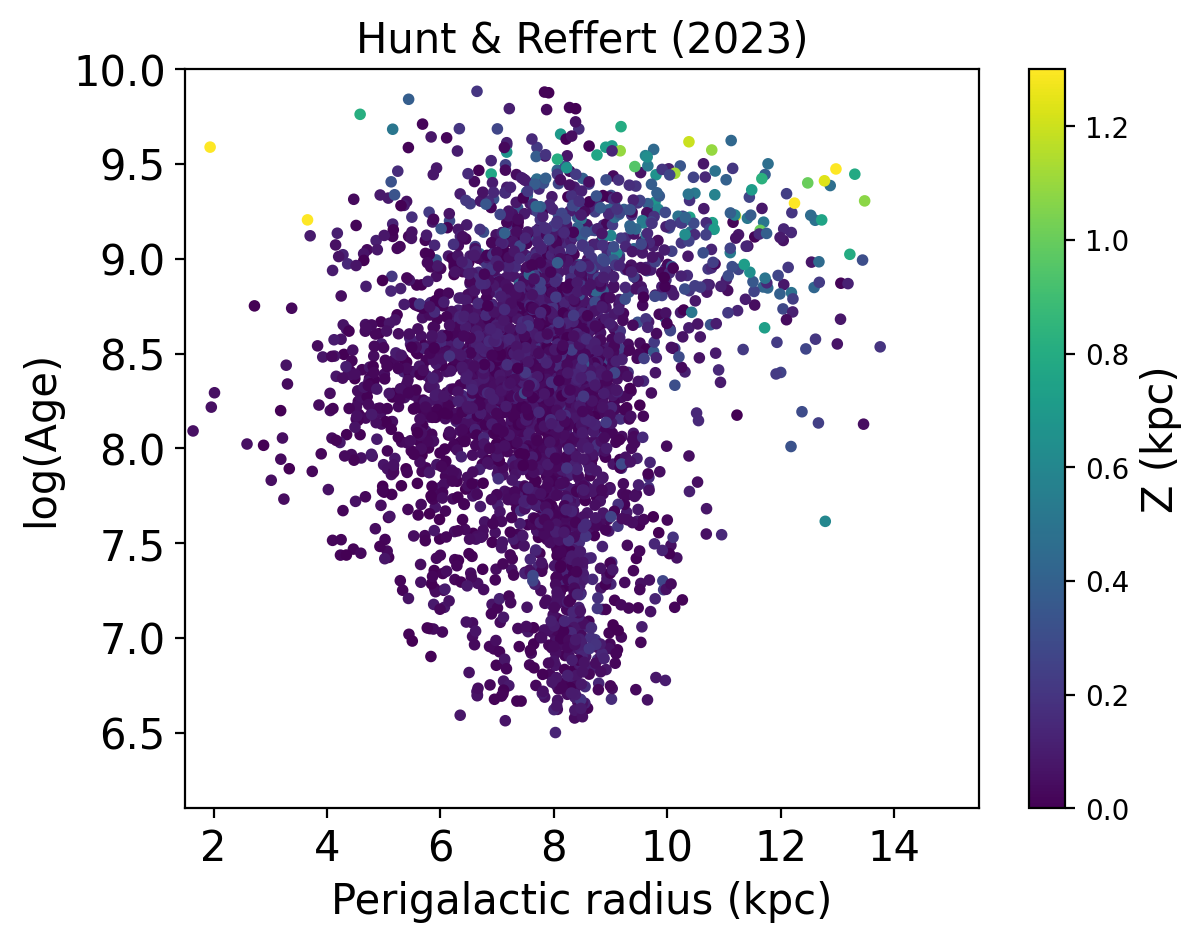}
    \caption{{\bf The perigalactic radius of the orbits of OCs in both catalogues as a function of their log(age) and distance from the Galactic disk.}}
    \label{fig:peri}
\end{figure*}

The sample studied here shows fewer older OCs inner to the solar circle and more outside, whereas the intermediate and young OCs are similarly distributed. This may point to the efficient destruction of older OCs in the inner regions, leading to better survival in the outer disk. 
{\bf This can be confirmed by Fig. \ref{fig:peri}, which shows the perigalactic radii of OCs in both catalogues as a function of their log(age) and height from the Galactic disk.
This figure shows that the older clusters have large perigalactic distances and OCs located at a higher distance from the Galactic disk seem to move with larger orbits.}

The larger values of $Z_{max}$ found for older OCs at all Galactic radii may point to a particular kinematic perturbation around 1 Gyr, resulting in either a heating of the disk or a kinematically decoupled population.

\section{Summary}\label{sec:summary}

In this study, we updated the OC catalogue by \cite{2020A&A...640A...1C} based on Gaia DR2 by incorporating the data from Gaia DR3. 
This catalogue consists of the fundamental properties of 1145 OCs, such as position, proper motion, radial velocity, distance, and age. 
This sample is accompanied by 3677 OCs from the catalogue presented by \cite{2023A&A...673A.114H}.
Using this extensive sample of 4006 OCs (816 in common), we studied their distribution and motion in the Galaxy. As the parameters of OCs differ between these catalogues, the analysis is presented for both samples separately.
We estimated their orbital parameters such as eccentricity, perigalactic and apogalactic distance and the maximum vertical height traced from the Galactic disk (Z$_{max}$) by deriving their orbits in the Galaxy. 
The main findings of this paper are summarised as follows:

\begin{enumerate}
    \item  The number distribution for OCs younger than 1 Gyr appears similar across the Galactocentric radial distance. In the case of OCs older than 1 Gyr, we note a radially extended distribution, with older OCs in the outer part of the disk.  Since older clusters show a wider distribution of vertical heights, a greater fraction of them can reach high galactic latitudes; therefore, they are less affected by interstellar extinction and can be observed up to larger distances.

    \item OCs younger than 1 Gyr have a very similar distribution of vertical height, whereas OCs older than 1 Gyr can reach significantly higher vertical heights. The maximum vertical height, $Z_{max}$, of the OCs is a function of age and Galactocentric radius. We derive relations for this dependency for the first time.

    \item The value of $Z_{max}$ is the lowest for young OCs (age $<$ 50 Myr - 263 OCs). The intermediate age ( 50 Myr $<$age $<$ 1 Gyr - 700 OCs) and the old cluster ($<$age $>$ 1 Gyr - 182 OCs) show larger values for the maximum vertical heights. We also note a distinct pattern in the $Z_{max}$ for the three age groups. We note that the two younger age groups show a similar pattern up to a radius of 9 kpc, beyond which the age group shows larger values for $Z_{max}$. The oldest age group also appears to have a differing pattern beyond 9 kpc, with the peak-like feature at a radius of 11 kpc. The pattern shown by the intermediate and the old age group of the OCs suggests a flaring of the disk beyond a radius of 9 kpc.

    \item For OCs younger than 10 Myr, median $Z_{max}$ is 63 pc; OCs in the age range 10 - 100 reach a median $Z_{max}$ of 69 pc; OCs in the 100 - 800 Myr reach a $Z_{max}$ of 100 pc and OCs older than 800 Myr have a median $Z_{max}$ of 221 pc. Therefore, we observe that the value of $Z_{max}$ increases as a function of the cluster's age.

    \item We note that at two radial distance values (8 kpcs and 9.5 kpc), the value of $Z_{max}$ jumps by nearly 20-30 pc. The confidence level of these two peaks is found to be 80 and 99\% respectively. This increased value of $Z_{max}$ has a width not more than 500 pc at the values mentioned above of the $R$.

    \item  The variation of $Z_{max}$ as a function of the $l$ suggests that the OCs between $l \sim 90$ and $l \sim 320$ have an increased $Z_{max}$. Also, the peak of the distribution tends to point to the third quadrant more than the second quadrant. We speculate that the higher $Z_{max}$ in these quadrants is related to the already identified warp in these regions of the Galactic disk.

    \item This study also shows that the OCs are involved in the diagonal ridges observed in the solar neighbourhood and are occupied mainly by the young OCs.

\end{enumerate}

\section*{Acknowledgements}
AS acknowledges the support from the SERB Power fellowship. We thank the anonymous referee for the valuable comments that enhance the quality of the paper.
This work has made use of data from the European Space Agency (ESA) mission
{\it Gaia} (\url{https://www.cosmos.esa.int/gaia}), processed by the {\it Gaia}
Data Processing and Analysis Consortium (DPAC,
\url{https://www.cosmos.esa.int/web/gaia/dpac/consortium}). Funding for the DPAC
has been provided by national institutions, in particular, the institutions
participating in the {\it Gaia} Multilateral Agreement.

%%%%%%%%%%%%%%%%%%%%%%%%%%%%%%%%%%%%%%%%%%%%%%%%%%
\section*{Data Availability}

We have used the publicly available Gaia DR3 catalogue, which can be accessed from:
\url{https://gea.esac.esa.int/archive/}

The complete electronic tables will be provided after the acceptance of the paper.

%%References section
\bibliography{Bibliography}
\end{document}